\documentclass[a4paper,11pt]{article}
\usepackage{jcappub} 
\usepackage{lineno}

\usepackage{xcolor}
\newcommand{\red}{\textcolor{black}}

\usepackage{subcaption}

\newcommand{\CosmoLattice}{{\tt ${\mathcal C}$osmo${\mathcal L}$attice}}

\arxivnumber{2512.21151} 

\title{\boldmath Acoustic gravitational waves from primordial curvature perturbations}

\author[a,b,c]{Zhuan Ning,}
\emailAdd{ningzhuan17@mails.ucas.ac.cn}

\author[f]{Zi-Yan Yuwen,}
\emailAdd{ziyan.yuwen@apctp.org}

\author[b,d]{Xiang-Xi Zeng,}
\emailAdd{zengxiangxi@sjtu.edu.cn}

\author[e]{Rong-Gen Cai,}
\emailAdd{caironggen@nbu.edu.cn}

\author[b,f]{Shao-Jiang Wang}
\emailAdd{schwang@itp.ac.cn}

\affiliation[a]{School of Fundamental Physics and Mathematical Sciences, Hangzhou Institute for Advanced Study (HIAS), University of Chinese Academy of Sciences (UCAS), Hangzhou 310024, China}
\affiliation[b]{Institute of Theoretical Physics, Chinese Academy of Sciences (CAS), Beijing 100190, China}
\affiliation[c]{University of Chinese Academy of Sciences (UCAS), Beijing 100049, China}
\affiliation[d]{School of Physical Sciences, University of Chinese Academy of Sciences (UCAS), Beijing 100049, China}
\affiliation[e]{Institute of Fundamental Physics and Quantum Technology \& School of Physical Science and Technology, Ningbo University, Ningbo, 315211, China}
\affiliation[f]{Asia Pacific Center for Theoretical Physics (APCTP), Pohang 37673, Korea}

\abstract{Standard perturbative calculations of scalar-induced gravitational waves (SIGWs) have neglected nonperturbative effects in the large-amplitude regime. We develop a hybrid numerical framework to signify nonperturbative effects on the stochastic gravitational wave (GW) background sourced by primordial curvature perturbations, focusing on the acoustic channel (fluid motions). Fully general-relativistic, spherically symmetric simulations are used to extract nonperturbative sound-shell profiles from isolated curvature peaks; these profiles are then embedded into three-dimensional lattice evolutions of relativistic hydrodynamics coupled to transverse-traceless metric perturbations to compute the acoustic GW spectra. The acoustic signal has a peak frequency determined by the comoving shell thickness, and its amplitude is extremely sensitive to the mean comoving separation of peaks, scaling approximately as $R_{*c}^{-7}$. We find a robust causal low-frequency tail $\propto k^{3}$, and the nonlinear hydrodynamic interactions can enhance the ultraviolet power. Comparing with SIGWs computed perturbatively from the same real-space configuration, we show that acoustic GWs can be amplified by an order of magnitude and display a peak shifted to a lower frequency in the large-amplitude regime. These results highlight the importance of nonperturbative effects for accurate predictions of stochastic GW signals induced from primordial curvature perturbations.}

\begin{document}
\maketitle
\flushbottom

\section{Introduction} \label{sec:introduction}

The inflationary paradigm~\cite{Starobinsky:1980te, Guth:1980zm, Linde:1981mu, Albrecht:1982wi} provides a compelling mechanism for generating primordial curvature perturbations that seed radiation relics and structure formations in the Universe. Quantum fluctuations of the inflaton are stretched to super-Hubble scales during inflation and become effectively classical curvature perturbations upon horizon exit. These perturbations later reenter the horizon during the radiation- and matter-dominated eras and can source a variety of observable phenomena (see, e.g., reviews~\cite{Achucarro:2022qrl, LISACosmologyWorkingGroup:2022jok, LISACosmologyWorkingGroup:2023njw} and references therein). At large scales, the curvature power spectrum is tightly constrained by observations of the cosmic microwave background (CMB)~\cite{Lewis:1999bs, Acquaviva:2002ud, Mollerach:2003nq, Planck:2018jri, Planck:2019kim, BICEP:2021xfz} and large-scale structure (LSS)~\cite{Tomita:1975kj, Matarrese:1993zf, Bernardeau:2001qr}. At much smaller scales, however, the power spectrum is largely unconstrained and may even be significantly enhanced in a variety of inflationary scenarios~\cite{Ivanov:1994pa, Kinney:1997ne, Leach:2000ea, Inoue:2001zt, Kinney:2005vj, Martin:2012pe, Alabidi:2012ex, Motohashi:2017kbs, Ballesteros:2017fsr, Germani:2017bcs, Garcia-Bellido:2017mdw, Bhaumik:2019tvl, Ballesteros:2020qam, Ragavendra:2020sop}. Enhanced small-scale curvature perturbations are of particular interest because they can simultaneously (i) produce primordial black holes (PBHs) via gravitational collapse of rare, large overdensities~\red{\cite{Zeldovich:1967lct, Hawking:1971ei, Carr:1974nx, Carr:1975qj, Chapline:1975ojl, Sasaki:2018dmp, Carr:2020gox, Carr:2026hot}}, and (ii) generate a stochastic background of scalar-induced gravitational waves (SIGWs) through nonlinear coupling between scalar and tensor modes~\red{\cite{Matarrese:1997ay, Acquaviva:2002ud, Ananda:2006af, Baumann:2007zm, Kohri:2018awv, Domenech:2021ztg}}. \red{More recent developments in PBH formation include studies of profile dependence, type-II configurations, nonspherical collapse, and PBH formation from negative-amplitude curvature perturbations~\cite{Musco:2018rwt, Uehara:2024yyp, Escriva:2025eqc, Ning:2026nfs, Escriva:2024lmm, Ning:2026jkk, Escriva:2026kov, Joana:2025gqf}. Subsequent developments in SIGWs have investigated non-Gaussian effects, higher-order effects, gauge issues, and lattice-based calculations~\cite{Espinosa:2018eve, Cai:2018dig, Bartolo:2018evs, Yuan:2019udt, Zhao:2026jax, Domenech:2025ccu, Zeng:2025cer, Caravano:2026hca}.} These observational channels, therefore, provide complementary probes of inflationary dynamics at small scales.

The standard calculation of SIGWs proceeds within the framework of cosmological perturbation theory: first-order scalar perturbations source second-order tensor perturbations, with both the scalar metric perturbations and the induced velocity perturbations of the background fluid contributing to the gravitational wave (GW) source. The resulting GW energy spectrum depends quadratically on the primordial curvature power spectrum for the Gaussian statistics. This perturbative formalism is valid as long as the relevant perturbations remain small. However, if perturbations reenter the horizon with amplitudes large enough to trigger nonperturbative dynamics, their nonlinear evolution can depart dramatically from perturbative expectations. Self-gravity effect can drive perturbed regions to decouple from the background and undergo substantial nonlinear evolution: rare but sufficiently large peaks may collapse to form PBHs\footnote{Recently, Ref.~\cite{Joana:2025gqf} pointed out that large negative curvature perturbations can also form PBHs. We do not consider this scenario here.}, while more abundant but subcritical peaks undergo a dynamical process of contraction, turnaround, and bounce. These highly nonlinear processes generically produce additional sources of gravitational radiation that are not captured by the standard SIGW framework, and they must be considered to obtain a complete characterization of GW signals associated with large-amplitude curvature perturbations. In particular, nonperturbative effects will modify the constraints on PBH abundances inferred from GW observations~\cite{Saito:2008jc, Saito:2009jt,Cai:2018dig}.

Several studies have investigated GW production beyond the perturbative SIGW calculation. Ref.~\cite{DeLuca:2019llr} considered GW generation from the non-spherical gravitational collapse of subcritical curvature peaks, although the amplitude of the resulting GW signal is typically small and merely measurable. For GW production during an early matter-dominated era, nonlinear dynamics of density perturbations on sub-Hubble scales, such as the formation, tilde interactions, and evaporation of matter halos, have been shown to amplify the GW signal to potentially observable levels~\cite{Assadullahi:2009nf, Jedamzik:2010hq, Eggemeier:2022gyo, Fernandez:2023ddy}. In Ref.~\cite{Fernandez:2023ddy}, the authors performed hybrid N-body and lattice simulations to study these nonlinear effects and reported a significant enhancement of the GW signal in the large-amplitude regime. During the radiation-dominated era, the competition between pressure gradients and self-gravity complicates the nonlinear dynamics. In our previous work~\cite{Ning:2025ogq}, we performed fully general-relativistic (GR), spherically symmetric (one-dimensional, 1D) simulations of the collapse of individual curvature peaks. We found that strong outward-propagating sound waves (``sound shells'') are excited in both subcritical, near-critical, and supercritical cases. The energy carried by these sound shells depends exponentially on the difference between the perturbation amplitude and the critical threshold for PBH formation, implying that nonperturbative effects can substantially amplify fluid motions. Although a single, spherically symmetric sound shell does not radiate GWs, the superposition and collision of sound waves from many peaks can produce a stochastic background of acoustic GWs. Therefore, nonlinear/nonperturbative enhancement of fluid motions may amplify the GW signals associated with primordial curvature perturbations.

Performing three-dimensional (3D), fully GR simulations that resolve PBH formation, the hydrodynamic response, and GW emission is numerically challenging. However, if the curvature perturbations can be modeled as a collection of sparse and discrete peaks~\cite{Iovino:2025xkq}, the sound waves generated by individual peaks will enter a regime of free propagation before collisions. In this case, one can extract the sound-shell profiles from 1D GR simulations and then use the sound shell model~\cite{Hindmarsh:2016lnk, Hindmarsh:2019phv, Guo:2020grp} to compute the GW signal produced by colliding sound waves. We applied this method in Ref.~\cite{Zeng:2025law} to semi-analytically estimate the GW spectrum and found that acoustic GWs can be observable in portions of the parameter space relevant for PBH formation, but under a simplified assumption of uniform spatial distribution of curvature peaks with identical amplitudes. Relaxing these assumptions and obtaining more robust predictions requires lattice simulations, which is the primary goal of the present paper. Moreover, since \red{the semi-analytical calculation of acoustic GWs in Ref.~\cite{Zeng:2025law}} neglects metric perturbations, it is important to compare the amplitude of acoustic GWs with that of SIGWs generated by the same curvature perturbations to assess the validity of this approximation. To this end, we perform hybrid simulations that combine 1D fully GR runs (to extract sound-shell profiles nonperturbatively) with 3D lattice evolutions of relativistic hydrodynamics and tensor perturbations (to compute the GW signal from an ensemble of curvature peaks). Then, we compare the resulting acoustic GW spectra with SIGW predictions and find that our acoustic GWs are essentially the SIGWs but manifesting the nonperturbative effects.

\red{The purpose of the present work is not to provide a model-specific prediction for a particular inflationary scenario, but to construct a controlled framework for quantifying nonperturbative effects in the acoustic channel of GW production from large primordial curvature perturbations. It is useful to distinguish three classes of inputs and assumptions. The physical framework consists of radiation domination, relativistic perfect-fluid dynamics, and primordial curvature peaks whose fully GR evolution generates outward sound shells. We then introduce benchmark assumptions and approximations to isolate the nonlinear physics: isolated spherical peaks, the Gaussian profile of Eq.~\eqref{eq:zeta}, a common characteristic scale $r_m$, initially a common amplitude $\mu$ before the distribution is generalized in Sec.~\ref{sec:amplitude_distribution}, random spatial placement with a non-overlap constraint, and the neglect of direct PBH influence and gravitational backreaction in the 3D stage. Finally, the comoving box size $L$, grid resolution $N$, and number of shells $N_s$ are numerical implementation parameters. Of these, $L$ and $N_s$ separately are simulation parameters, while their combination $R_{*c}=L/N_s^{1/3}$ is the physical ensemble parameter controlling the shell abundance once finite-volume effects are controlled. In a concrete cosmological model, the primordial spectrum and statistics would determine the distributions of $\mu$, $r_m$, profile shape, and spatial correlations, which could then be passed through the framework developed here.}

\begin{figure}[htbp]
    \centering
    \includegraphics[width=0.9\linewidth]{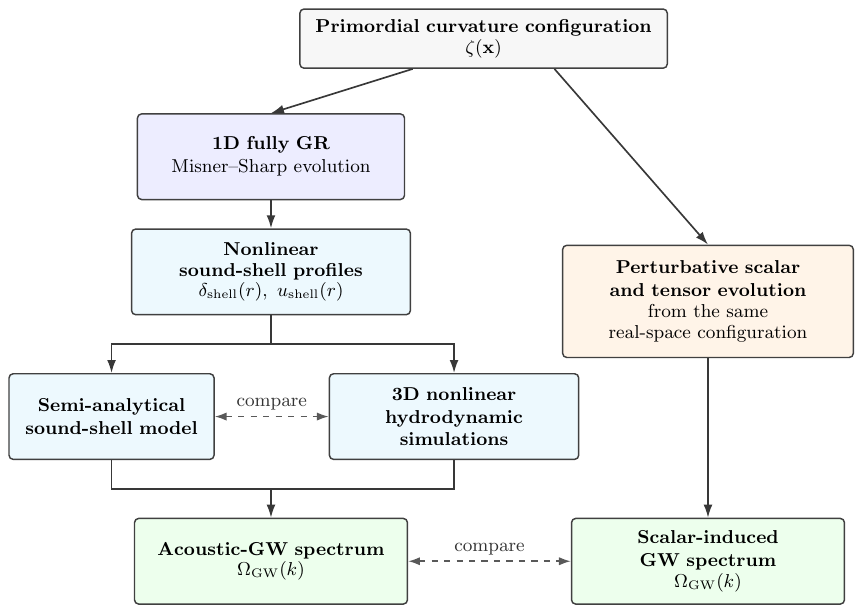}
    \caption{\red{Overview of the calculation. The primordial curvature configuration is evolved along a fully relativistic branch to obtain nonlinear sound-shell profiles, which provide the inputs to the semi-analytical and 3D acoustic-GW calculations. In parallel, the same underlying real-space curvature configuration is used for the perturbative SIGW calculation. Dashed double-headed arrows denote direct comparisons.}}
    \label{fig:workflow_overview}
\end{figure}

\red{The flowchart in Fig.~\ref{fig:workflow_overview} makes explicit that the acoustic-GW calculation starts from the primordial curvature perturbation, which is evolved with the 1D fully GR Misner--Sharp system to obtain the nonlinear sound-shell profiles. These profiles are then used as inputs to both the semi-analytical sound-shell model and the 3D nonlinear hydrodynamic simulations. In parallel, the same underlying primordial real-space curvature configuration provides the starting point for the perturbative SIGW calculation. The resulting acoustic-GW and SIGW spectra are then compared.}

This paper is organized as follows: in Sec.~\ref{sec:SW}, we briefly review the 1D fully GR simulation results of Ref.~\cite{Ning:2025ogq} and compare the nonperturbative sound-shell profiles with linear perturbation theory to highlight nonperturbative effects. In Sec.~\ref{sec:semi}, we present the general formalism for GW production from freely propagating sound waves, deriving the fluid velocity field and its power spectrum within the sound shell model and relating these quantities to the GW spectrum. Sec.~\ref{sec:setup} describes our 3D lattice simulation setup, including the equations of motion and numerical methods. In Sec.~\ref{sec:results}, we show our numerical results and compare them with our semi-analytical estimates~\cite{Zeng:2025law} and with SIGW predictions. Finally, Sec.~\ref{sec:conclusions} summarizes our conclusions and discusses implications. Throughout this paper, we adopt natural units with $c = \hbar = 8\pi G = 1$, so that the reduced Planck mass is $M_\mathrm{Pl} =1/\sqrt{8\pi G}= 1$.

\section{Sound waves from collapsing peaks in curvature perturbations} \label{sec:SW}

In this section, we will first use the fully GR simulation within the Misner--Sharp formalism in Sec.~\ref{subsec:MS} to nonperturbatively extract the sound-wave density and velocity profiles around the collapsing peak in curvature perturbations. As a comparison, we will then use the traditional linear perturbation theory in Sec.~\ref{subsec:compare_SW} to calculate the same sound-wave profiles. The small-perturbation case serves as a cross-check of our Misner--Sharp simulations. However, the large-perturbation case from Misner--Sharp simulations exhibits a significant nonperturbative effect that the traditional linear perturbative treatment cannot capture, which will be crucial to understand our acoustic GWs when compared to the SIGW within the traditional perturbative treatment. The obtained sound-wave profiles will be used as inputs for later semi-analytical sound-shell calculations (Sec.~\ref{sec:semi}) and 3D real-space lattice simulations (Sec.~\ref{sec:setup}) of the induced acoustic GWs (Sec.~\ref{sec:results}).

\subsection{Fully GR simulations}\label{subsec:MS}

As described in Ref.~\cite{Ning:2025ogq}, we perform a fully GR, spherically symmetric simulation of the individual collapse of the peak in curvature perturbation using the Misner--Sharp formalism~\cite{Misner:1964je} and then extract the resulting sound-shell profile. In this formalism, the metric ansatz is
\begin{equation}
    \mathrm{d}s^2 = -A(t,r)^2\mathrm{d}t^2 + B(t,r)^2\mathrm{d}r^2 + R(t,r)^2\mathrm{d}\Omega^2,
\end{equation}
where the metric components $A$, $B$, and $R$ are functions of the cosmic time $t$ (in the absence of perturbations) and the comoving radial coordinate $r$, and $\mathrm{d}\Omega^2 = \mathrm{d}\theta^2+\sin^2\theta \mathrm{d}\varphi^2$ is the unit-sphere line element. The cosmic fluid is taken to be a perfect fluid with an equation of state $p = \omega\rho$, where $p$ and $\rho$ denote the pressure and energy density, respectively.

At the initial time, the perturbation length scale is much larger than the Hubble radius; thus, we set initial conditions using the long-wavelength approximation~\cite{Salopek:1990jq, Polnarev:2006aa, Polnarev:2012bi, Harada:2015yda}. To facilitate comparison with cosmological perturbation theory, we encode the curvature perturbation directly as a multiplicative perturbation in the scale factor rather than by introducing a non-homogeneous curvature $K(r)$ into the spatial part of the Friedmann-Lema\^{i}tre-Robertson-Walker (FLRW) metric. Thus, the line element for the initial curvature perturbation reads
\begin{equation}
    \mathrm{d}s^2 = -\mathrm{d}t^2 + a(t)^2e^{2\zeta(r)}\left(\mathrm{d}r^2 + r^2\mathrm{d}\Omega^2\right),
\end{equation}
where $a(t)$ is the scale factor and $\zeta(r)$ is the super-horizon comoving curvature perturbation~\cite{Musco:2018rwt}. In this work, we adopt a Gaussian curvature profile,
\begin{equation} \label{eq:zeta}
    \zeta(r) = \mu e^{-(r/r_m)^2},
\end{equation}
where $r_m$ is the initial perturbation length scale and $\mu$ is the amplitude parameter. \red{The Gaussian profile in Eq.~\eqref{eq:zeta} is not intended as a prediction of a particular inflationary model, nor do our conclusions require primordial curvature peaks to have exactly this shape. We adopt it as a smooth, localized, single-scale benchmark that enables a controlled scan across the perturbative, nonlinear, near-critical, and PBH-forming regimes by varying a single amplitude parameter $\mu$.}

The relevant conceptual point is that our framework constitutes a mapping
\begin{align}
    \zeta(r)
    \longrightarrow
    \{\delta_{\rm shell}(r),u_{\rm shell}(r)\}_{\rm GR}
    \longrightarrow
    P_v(k)
    \longrightarrow
    \Omega_{\rm GW}(k),
\end{align}
so that changing the primordial profile modifies the final GW spectrum through the nonperturbatively generated sound-shell profile rather than invalidating the framework itself.

The dependence on the initial profile has been investigated directly in our accompanying fully relativistic study, Ref.~\cite{Ning:2025ogq}, where generalized Gaussian and smoothed top-hat curvature profiles were evolved. Different profiles lead to quantitatively different PBH thresholds, shell thicknesses, and shell energies, while the formation and outward propagation of sound shells persist across the profiles investigated. In particular, the profile steepness appreciably affects the resulting shell properties. Since the shell thickness controls the characteristic acoustic-GW scale and the shell profile determines the fluid velocity power spectrum, the precise GW peak position, normalization, and detailed spectral shape obtained below should therefore be regarded as profile dependent.

For simplicity, we further assume that all curvature peaks share the same characteristic scale $r_m$. \red{This is representative of a narrow, approximately monochromatic primordial spectrum centered around a characteristic scale and should be understood as a simplifying benchmark rather than a generic prediction. A common $r_m$ fixes a common primordial length scale and approximately the same horizon-entry epoch, thereby setting the overall characteristic scale of the resulting sound-shell thickness. The actual shell thickness is nevertheless modified by the subsequent nonlinear evolution and can depend appreciably on the peak amplitude $\mu$, particularly near and above the collapse threshold. This choice allows us to isolate the nonperturbative dependence on peak amplitude and shell dynamics without simultaneously introducing a distribution of formation times and length scales.}

Concretely, we choose $r_m = 10r_H$ for demonstration, where $r_H$ is the comoving Hubble radius at the initial time of the 1D simulation. With this choice, the perturbation crosses the horizon at $t_m = 50r_H$ in the radiation-dominated era ($\omega = 1/3$).

For a fixed shape of the initial curvature perturbation, $\mu$ controls its amplitude. The critical value $\mu_c$ separates two qualitatively different outcomes: for $\mu<\mu_c$ no apparent horizon forms, whereas for $\mu>\mu_c$ an apparent horizon forms and a PBH is produced. The critical value is not an analytic input. We determine $\mu_c$ numerically by bracketing subcritical and supercritical solutions and successively narrowing the interval in $\mu$. PBH formation is diagnosed by the formation of an apparent horizon, corresponding to $2M/R=1$ in the Misner--Sharp variables during the collapsing evolution. For the Gaussian profile in Eq.~\eqref{eq:zeta}, we obtain $\mu_c\simeq0.79579\pm10^{-5}$, where the quoted uncertainty is the numerical bracketing accuracy. Further details of the numerical implementation are given in Ref.~\cite{Ning:2025ogq}.

In the Misner--Sharp description, $U\equiv D_tR$ describes the radial motion of the areal radius and contains the homogeneous Hubble expansion. Its FLRW background value is $\bar U=HR$. The quantity relevant for the outgoing acoustic perturbation is therefore the peculiar radial velocity, $u\equiv U-\bar U$, which is mapped to the comoving radial velocity $\mathrm{d}r/\mathrm{d}\eta$ in the asymptotic FLRW region.

\begin{figure}[htbp]
    \centering
    \includegraphics[width=\linewidth]{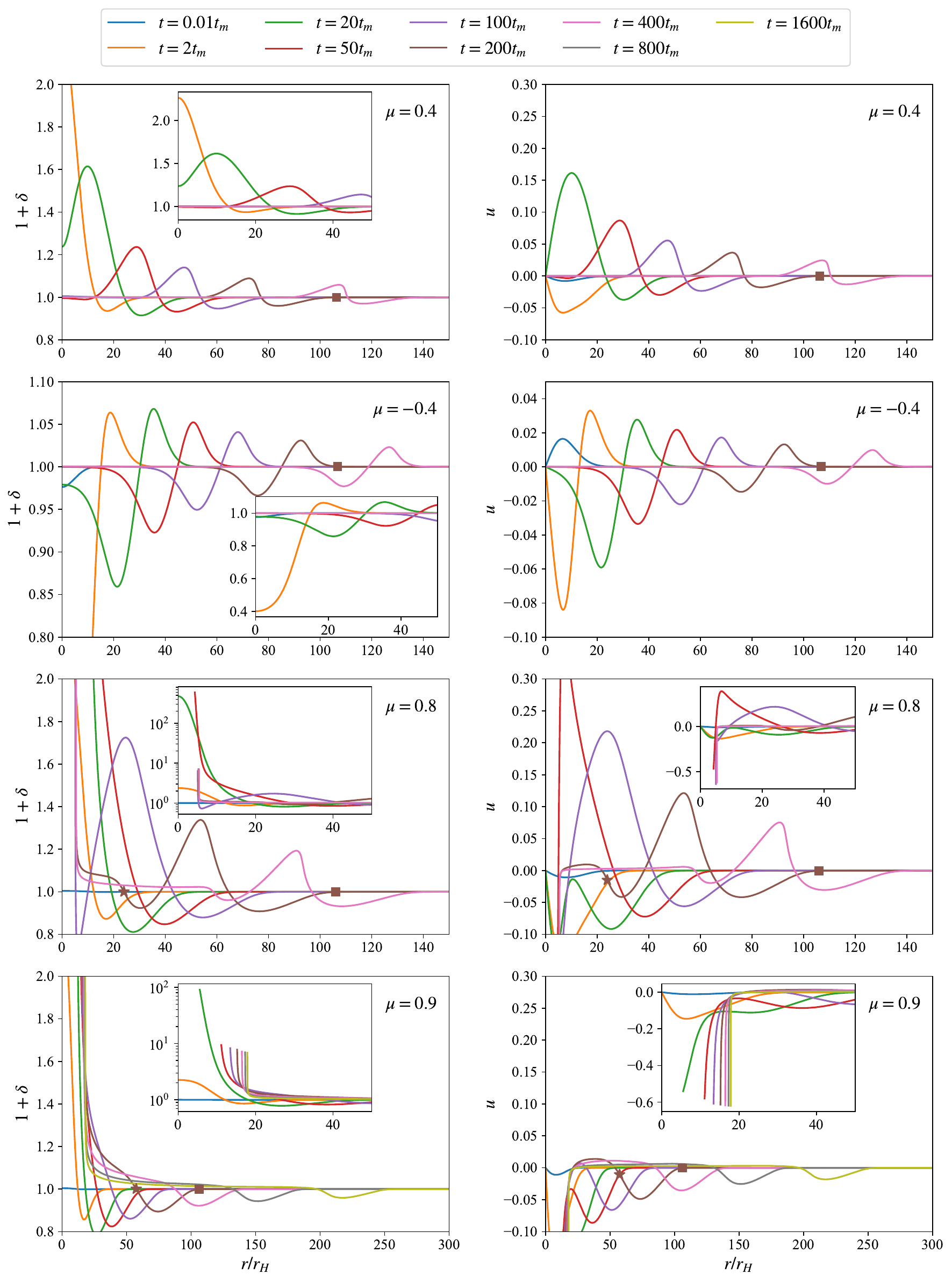}
    \caption{\red{Normalized density $1+\delta=\rho/\bar\rho$ (left column) and radial velocity $u$ (right column) at several time slices for the subcritical ($\mu = 0.4$), negative-amplitude ($\mu = -0.4$), near-critical ($\mu = 0.8$), and supercritical ($\mu = 0.9$) perturbations from top to bottom. The main panels use zoomed vertical ranges to resolve the outgoing sound-shell structures, while the insets display the full vertical extent of the affected central curves. At $t=200t_m$, stars mark the inner edges of the profiles retained for the 3D initial conditions for $\mu=0.8$ and $0.9$, and squares mark the outer edges for all cases. For the noncollapsing cases, the profiles are retained from the origin to the square; for $\mu=0.8$ and $0.9$, the region between the star and square is retained.}}
    \label{fig:shell_profile}
\end{figure}

Figure~\ref{fig:shell_profile} displays the normalized density $1+\delta=\rho/\bar\rho$ in the left column and the radial velocity $u$ in the right column at several time slices for four representative cases: subcritical ($\mu=0.4$), negative-amplitude ($\mu=-0.4$), near-critical ($\mu=0.8$), and supercritical ($\mu=0.9$). The main panels use zoomed vertical ranges to resolve the outgoing sound-shell structures. Insets in the affected panels show the full vertical extent of the previously clipped central large-amplitude curves. The density insets for $\mu=0.8$ and $0.9$ use logarithmic vertical axes to accommodate the large central overdensities, while those for $\mu=0.4$ and $-0.4$ use linear axes.

For the subcritical and near-critical configurations, pressure gradients drive outward fluid motion and produce both overdense and underdense shell structures. The negative-amplitude perturbation yields an outer overdense shell and an inner underdense shell. Near the collapse threshold, self-gravity has already driven the system deeply into the nonlinear regime while pressure gradients remain capable of producing a pronounced outward rebound, giving the strongest sound shell among the cases considered here. In the supercritical case, collapse dominates the central dynamics and only a broad underdense outgoing component remains.

These nonperturbatively generated density and velocity profiles provide the initial data for the second stage of our calculation: the freely propagating sound-shell components are subsequently used both in the semi-analytical sound-shell calculation and in the 3D lattice simulations of acoustic-GW production. To quantify the differences between the cases, we characterize the outgoing sound shell by its peak density amplitude
\begin{align}
    A_\delta(t)=\max_{r\in\mathcal S(t)}|\delta(t,r)|,
\end{align}
and by its characteristic shell thickness $d$. The retained shell region $\mathcal S(t)$ and the corresponding diagnostics are defined in Sec.~\ref{subsec:compare_SW} and summarized in Table~\ref{tab:shell_diagnostics}.

\subsection{Comparison with linear evolution} \label{subsec:compare_SW}

In the traditional cosmological perturbation theory, if we only consider the first-order scalar perturbations, the metric can be written as~\cite{Kodama:1984ziu, Mukhanov:1990me, Malik:2008im}
\begin{equation}
    \mathrm{d}s^2 = a(\eta)^2\left[-(1+2\phi)\mathrm{d}\eta^2 + 2\partial_i\beta\mathrm{d}\eta\mathrm{d}x^i + \left[(1-2\psi)\delta_{ij} + 2\partial_i\partial_j E\right]\mathrm{d}x^i\mathrm{d}x^j\right],
\end{equation}
where the scalar metric perturbations $\phi$, $\beta$, $\psi$, and $E$ are functions of $\eta$ and spatial coordinates $\vec{x}$. The comoving three-velocity of the fluid is expressed in terms of the velocity potential $v$ as $v^i = \delta^{ij}\partial_j v$.
The first-order Einstein and energy-momentum conservation equations yield
\begin{subequations}
    \begin{align}
        &3\mathcal{H}(\psi' + \mathcal{H}\phi) - \nabla^2[\psi + \mathcal{H}\sigma] = -\frac{a^2\delta\rho}{2M_\mathrm{Pl}^2}, \\
        &\psi' + \mathcal{H}\phi = -\frac{1}{2M_\mathrm{Pl}^2}a^2(\bar{\rho} + \bar{p})(v+\beta), \\
        &\delta\rho' + 3\mathcal{H}(\delta\rho + \delta p) - 3(\bar{\rho} + \bar{p})\psi' + (\bar{\rho} + \bar{p})\nabla^2(v+\beta+\sigma) = 0, \\
        &v' + \beta' + (1-3c_s^2)\mathcal{H}(v+\beta) + \phi + \frac{\delta p}{\bar{\rho} + \bar{p}} = 0,
    \end{align}
\end{subequations}
where $\mathcal{H} \equiv a'/a$ is the conformal Hubble parameter, primes denote derivatives with respect to conformal time, $\bar{\rho}$ and $\bar{p}$ are the background energy density and pressure, respectively, $c_s^2 \equiv \mathrm{d}p/\mathrm{d}\rho = \omega$ is the sound speed, and $\sigma \equiv E' - \beta$. 

To compare with the Misner--Sharp formalism, we take the comoving orthogonal gauge ($\beta = v = 0$) and impose spherical symmetry (so the Laplacian operator reduces to $\nabla^2 = r^{-2}\partial_r(r^2\partial_r)$). Hence, one can derive the evolution equations for $\psi$ as
\begin{equation}
    \psi'' + \left(\frac{5+3\omega}{2}\mathcal{H}+\frac{\mathcal{H}'}{\mathcal{H}}\right)\psi' - \omega\nabla^2\psi = 0.
\end{equation}
Moreover, $\phi$ and $\delta\rho$ can be expressed in terms of $\psi$ as
\begin{equation}
    \phi = -\frac{\psi'}{\mathcal{H}}, \quad \frac{\delta\rho}{\bar{\rho}} = -\frac{1+\omega}{\omega}\phi.
\end{equation}
Transforming to cosmic time $t$ via $\mathrm{d}t = a\mathrm{d}\eta$, these relations become
\begin{align}
    &\ddot{\psi} + \left(\frac{3(3+\omega)}{2}H+\frac{\dot{H}}{H}\right)\dot{\psi} - \frac{\omega}{a^2}\nabla^2\psi = 0, \\
    &\phi = -\frac{\dot{\psi}}{H}, \quad \frac{\delta\rho}{\bar{\rho}} = -\frac{1+\omega}{\omega}\phi,
\end{align}
where $H \equiv \dot{a}/a$ is the Hubble parameter and dots denote derivatives with respect to the cosmic time. The initial condition for $\psi$ is set by the comoving curvature perturbation $\zeta$ as
\begin{equation}
    \zeta \equiv -\psi + \mathcal{H}(v+\beta) = -\psi,
\end{equation}
and $\dot{\psi}$ follows from $\dot{\psi} = -H\phi$. The metric perturbation $\phi$ at the initial time can be related to Misner--Sharp variables by 
\begin{align}
\phi = (A^2-1)/2.
\end{align}
When perturbations are small, the fully nonperturbative Misner--Sharp formalism should reproduce the results of linear perturbative evolution, which can serve as a verification of our correspondence of quantities between $(A, \zeta)$ and $(\phi, \psi)$ in the Misner--Sharp simulation and linear perturbative evolution, respectively.

\begin{figure}[htbp]
    \centering
    \includegraphics[width=\linewidth]{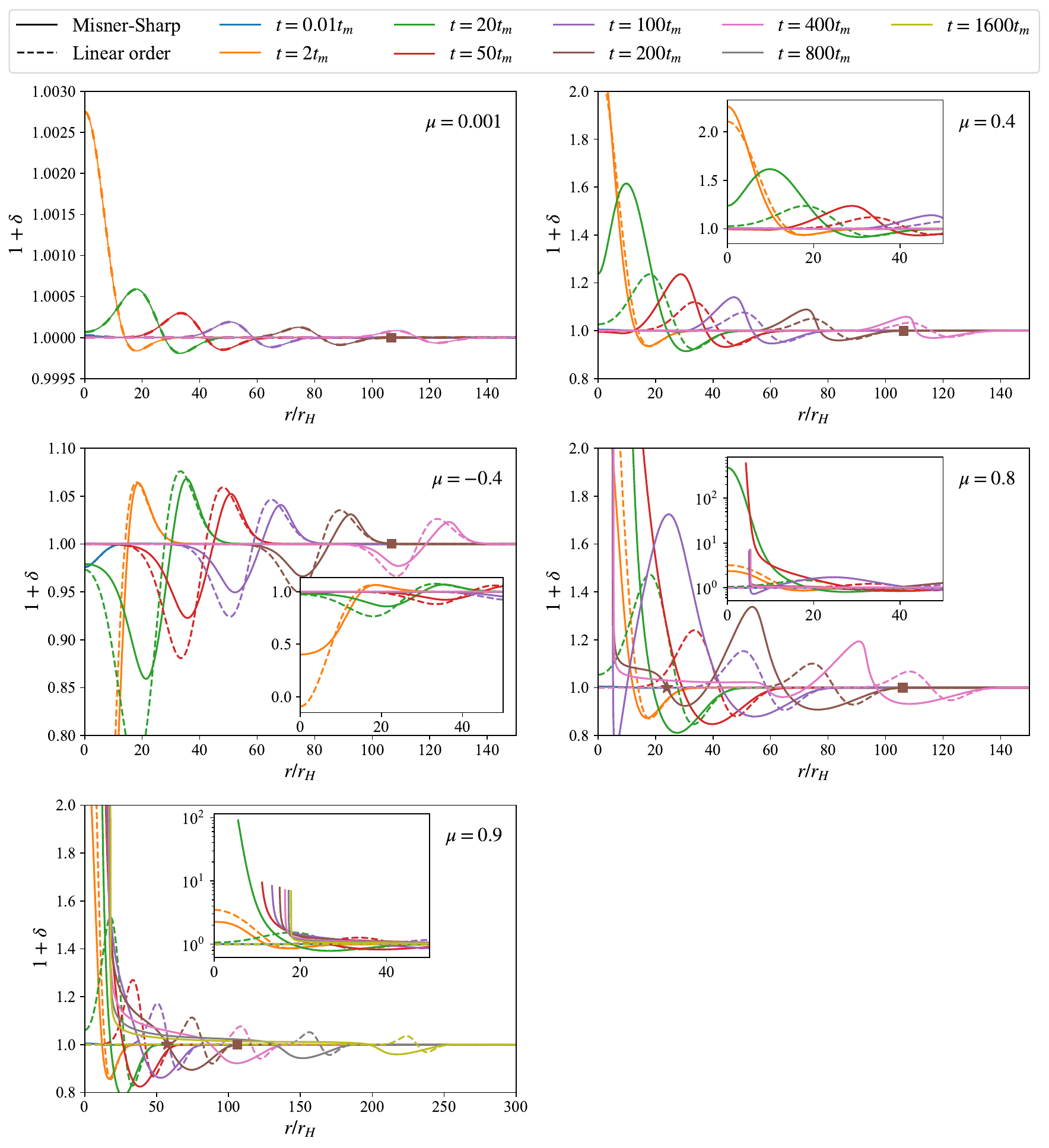}
    \caption{\red{Comparison of the normalized-density profiles $1+\delta=\rho/\bar\rho$ from the fully nonlinear Misner--Sharp simulations (solid lines) and linear evolution (dashed lines) at several time slices for $\mu=0.001$, $0.4$, $-0.4$, $0.8$, and $0.9$. For $\mu=0.001$, the solid and dashed lines coincide. At $t=200t_m$, stars mark the inner edges of the retained intervals for $\mu=0.8$ and $0.9$, and squares mark the outer edges for all cases.}}
    \label{fig:compare_SW}
\end{figure}

\red{Figure~\ref{fig:compare_SW} compares the normalized-density profiles from the fully nonlinear Misner--Sharp simulations (solid lines) with the linear perturbative evolution (dashed lines) at several time slices for $\mu = 0.001$, $0.4$, $-0.4$, $0.8$, and $0.9$. For the small-amplitude case, the two solutions coincide, as expected. For $\mu = 0.4$, the Misner--Sharp solution produces stronger sound waves than the linear prediction, whereas for $\mu = -0.4$ it yields a weaker sound wave. The near-critical result exhibits a much larger enhancement and a broader shape than the linear solution. For $\mu = 0.9$, the absence of an overdense outgoing shell in the Misner--Sharp simulation produces a markedly different profile relative to linear theory. These differences directly affect the resulting GW signal.}

Because the profiles of $\delta$ cross zero, a pointwise relative difference such as $(\delta_{\rm GR}-\delta_{\rm lin})/\delta_{\rm lin}$ becomes singular near zero crossings and can therefore be misleading. We instead define the volume-weighted relative profile difference
\begin{align}
    \Delta_\delta(t)
    =\left[
    \frac{\displaystyle\int_{\mathcal S(t)}\mathrm{d}r\,r^2
    |\delta_{\rm GR}(t,r)-\delta_{\rm lin}(t,r)|^2}
    {\displaystyle\int_{\mathcal S(t)}\mathrm{d}r\,r^2
    |\delta_{\rm lin}(t,r)|^2}
    \right]^{1/2},
\end{align}
and the GR-to-linear peak-amplitude ratio
\begin{align}
    \mathcal R_A(t)=\frac{A_{\delta,\rm GR}(t)}{A_{\delta,\rm lin}(t)}.
\end{align}
Here $\mathcal S=[r_{\rm in},r_{\rm out}]$ is selected using the same prescription as the 3D initial-data construction. For the noncollapsing cases, $r_{\rm in}=0$. For $\mu=0.8$ and $0.9$, $r_{\rm in}$ is the first zero crossing of the GR density contrast, so that the central strong-gravity region surrounding the PBH is excluded. The outer edge $r_{\rm out}$ is the first point outside the outer lobe satisfying
\begin{align}
    |\delta_{\rm GR}|<0.01\max(-\delta_{\rm GR}).
\end{align}
Stars in Figs.~\ref{fig:shell_profile} and~\ref{fig:compare_SW} mark the inner edges for $\mu=0.8$ and $0.9$, while squares mark the outer edges for all cases. For the noncollapsing profiles, the interval from the origin to the square is retained; for $\mu=0.8$ and $0.9$, the central region inside the star is replaced by the homogeneous background with zero peculiar velocity.

For a freely propagating spherical sound wave in a radiation fluid, Ref.~\cite{Ning:2025ogq} gives
\begin{align}
    u\equiv U-\bar U\simeq\frac{\delta}{4c_s},
    \qquad c_s=\frac{1}{\sqrt{3}},
\end{align}
and verifies this relation numerically in the sound-shell region. The density and velocity profiles therefore carry the same leading profile information in this region up to an approximately constant factor. We consequently report the density-based diagnostics above, while retaining the velocity panels in Fig.~\ref{fig:shell_profile} as a direct visual check of the fluid motion.

We define $d$ uniformly as the width of the outermost shell. Its inner edge is the last density zero crossing before $r_{\rm out}$ and its outer edge is $r_{\rm out}$. This outermost shell is underdense for positive $\mu$ and overdense for $\mu=-0.4$. All diagnostics are evaluated at the common reference time $t=200t_m$. The GR and linear profiles are compared on the same radial grid without shifting their peaks. The results are summarized in Table~\ref{tab:shell_diagnostics}.

\begin{table}[htbp]
    \centering
    \caption{\red{Density-profile diagnostics at the common reference time $t=200t_m$. Here $A_{\delta,\rm GR}$ is the maximum absolute GR density contrast in the retained interval $\mathcal S$, $d$ is the width of the outermost shell, $\mathcal R_A$ is the GR-to-linear peak-amplitude ratio, and $\Delta_\delta$ is the volume-weighted relative profile difference.}}
    \label{tab:shell_diagnostics}
    \begin{tabular}{c c c c c c}
        \hline\hline
        $\mu$ & $t/t_m$ & $A_{\delta,\rm GR}$ & $d/r_H$ & $\mathcal R_A$ & $\Delta_\delta$ \\
        \hline
        $0.001$ & $200$ & $1.25\times10^{-4}$ & $24.4$ & $1.001$ & $0.00161$ \\
        $0.4$   & $200$ & $0.0888$             & $29.2$ & $1.769$ & $0.987$ \\
        $-0.4$  & $200$ & $0.0337$             & $21.7$ & $0.676$ & $0.446$ \\
        $0.8$   & $200$ & $0.338$              & $41.6$ & $3.361$ & $2.556$ \\
        $0.9$   & $200$ & $0.106$              & $48.7$ & $0.935$ & $1.729$ \\
        \hline\hline
    \end{tabular}
\end{table}

\red{For $\mu=0.001$, $\Delta_\delta\simeq0.00161$, quantitatively confirming agreement with linear theory at the $0.16\%$ level. For $\mu=0.4$, the peak amplitude is enhanced by about $76.9\%$, whereas for $\mu=-0.4$ it is suppressed by about $32.4\%$. The near-critical case gives $\mathcal R_A=3.361$ and $\Delta_\delta=2.556$, demonstrating both strong enhancement and profile deformation. For $\mu=0.9$, the peak amplitude is suppressed by about $6.5\%$, while $\Delta_\delta=1.729$, reflecting the substantial change in the morphology of the outgoing profile. Changing the outer-tail threshold from $1\%$ to $0.5\%$ or $2\%$ changes $\Delta_\delta$ by less than $5.1\times10^{-5}$, while shifting the inner boundary by $\pm2r_H$ in the two PBH-forming cases changes it by less than $3.4\times10^{-4}$.}

\section{Semi-analytical calculations of acoustic gravitational waves} \label{sec:semi}

In this section, we will adopt the original sound shell model~\cite{Hindmarsh:2016lnk, Hindmarsh:2019phv, Guo:2020grp} to outline a semi-analytical procedure to compute the GW energy spectrum produced by an ensemble of freely propagating sound waves. Only in our case, these sound waves are generated by collapsing curvature perturbations instead of expanding bubble walls; hence, the recent progress in improving the sound shell model~\cite{Cai:2023guc,RoperPol:2023dzg,Sharma:2023mao} (when the lifetime of bubble walls is comparable to the lifetime of sound waves) is not considered here since there is no bubble wall at all in our case.

\subsection{GW energy spectrum}

Neglecting subdominant scalar and vector metric perturbations\footnote{Ref.~\cite{Giombi:2025tkv} has studied the effect of scalar perturbations on the GW spectrum in the context of cosmological first-order phase transition.}, the metric of a spatially flat FLRW spacetime with tensor perturbations reads
\begin{align}
    \mathrm{d}s^2 = -\mathrm{d}t^2 + a(t)^2\left[\delta_{ij} + h_{ij}(t, \vec{x})\right] \mathrm{d}x^i\mathrm{d}x^j,
\end{align}
where $h_{ij}$ denotes the transverse traceless (TT) tensor perturbation. In comoving Fourier space, the tensor perturbations are sourced by the TT part of the energy-momentum tensor $T_{ij}^{\mathrm{TT}}\equiv a^2\pi^{\mathrm{TT}}_{ij}$, through the perturbed Einstein equation
\begin{align}\label{eq:eom}
    h_{ij}^{\prime\prime}(t, \vec{k}) + 2\mathcal{H}h_{ij}^{\prime}(t, \vec{k}) + k^2h_{ij}(t, \vec{k}) = \frac{2}{M_\mathrm{Pl}^2}a^2\pi_{ij}^{\mathrm{TT}}(t, \vec{k}).
\end{align}
The TT projection is performed with the operator $\Lambda_{ijlm} \equiv \frac{1}{2}(P_{il}P_{jm} + P_{im}P_{jl} - P_{ij}P_{lm})$, where $P_{ij}(\hat{k}) \equiv \delta_{ij} - \hat{k}_i\hat{k}_j$ and $\hat{k}_{i} \equiv k_{i}/k$. The GW energy density $\rho_{\mathrm{GW}}$ is defined as
\begin{align}
    \rho_{\mathrm{GW}}(t) = \frac{M_\mathrm{Pl}^2}{4}\langle \dot{h}_{ij}(t, \vec{x})\dot{h}_{ij}(t, \vec{x}) \rangle.
\end{align}
One can define the power spectrum of $\dot{h}_{ij}$ as
\begin{align}
    \langle \dot{h}_{ij}(t, \vec{k})\dot{h}_{ij}(t, \vec{q}) \rangle = (2\pi)^3 \delta^3(\vec{k} + \vec{q}) P_{\dot{h}}(t, k).
\end{align}
In cosmology, it is conventional to define the dimensionless energy density fraction of the GWs as
\begin{align}
    \Omega_{\mathrm{GW}}(t) \equiv \frac{\rho_{\mathrm{GW}}(t)}{\rho_{c}(t)},
\end{align}
where $\rho_{c}(t)=3M_\mathrm{Pl}^2H^2$ is the critical energy density at time $t$, then the dimensionless GW energy spectrum follows
\begin{align}\label{eq:GW_spctrum}
\Omega_{\mathrm{GW}}(t,k)\equiv \frac{\mathrm{d}\Omega_{\mathrm{GW}}(t) }{\mathrm{d}\ln k} =\frac{1}{\rho_c(t) }\frac{\mathrm{d}\rho_{\mathrm{GW}}(t)}{\mathrm{d}\ln k} =
    \frac{k^3}{24\pi^2 H^2}P_{\dot{h}}(t, k)
    =\frac{k^3}{24\pi^2 H^2a^2}P_{h^{\prime}}(\eta, k),
\end{align}
where $P_{h'}$ is the power spectrum of $h_{ij}'$. To solve Eq.~\eqref{eq:eom}, it is common to use the Green function method. After introducing a Green function $G(x, x_0)$ with the following boundary conditions,
\begin{align}
    G(x\leq x_0) = 0, ~~~~~\left. \frac{\partial G(x,x_0)}{\partial x}\right|_{x=x_0^{+}} =1,
\end{align}
the formal solution of $h_{ij}$ is
\begin{align}
    h_{ij}(t, \vec{k}) = \frac{2}{M_\mathrm{Pl}^2}\int_{x_0}^{x}\mathrm{d}x_1 G(x,x_1)\frac{a^{2}(\eta_1)\pi^{\mathrm{TT}}_{ij}(\eta_1, \vec{k})}{k^2},
\end{align}
where $x = k\eta$ and $x_0$ is the time when the sound waves begin to propagate freely.
From the above solution, one can obtain the two-point correlation function of $h_{ij}'$ in terms of the unequal time correlator (UETC) of the source,
\begin{align}
    \langle h^{\prime}_{ij}(\eta, \vec{k})h^{\prime}_{ij}(\eta, \vec{q}) \rangle =& \left(\frac{2}{M_\mathrm{Pl}^2}\right)^2\int_{x_0}^{x}\mathrm{d}x_1\int_{x_0}^{x}\mathrm{d}x_2 \frac{\partial G(x,x_1)}{\partial x}\frac{\partial G(x,x_2)}{\partial x}\nonumber\\ &\times\frac{a^2(\eta_1)a^2(\eta_2)}{q^2}\langle \pi^{\mathrm{TT}}_{ij}(\eta_1,\vec{k})\pi^{\mathrm{TT}}_{ij}(\eta_2,\vec{q}) \rangle.
\end{align}
Therefore, the key step is to evaluate the UETC of $\pi_{ij}^{\mathrm{TT}}$. Assuming spatial homogeneity of the Universe, the UETC can be parameterized as~\cite{Hindmarsh:2015qta}
\begin{align}\label{eq:UETC}
    \langle \pi^{\mathrm{TT}}_{ij}(\eta_1,\vec{k})\pi^{\mathrm{TT}}_{ij}(\eta_2,\vec{q}) \rangle = \Pi^2(k, \eta_1, \eta_2)(2\pi)^2\delta^3(\vec{k} + \vec{q}).
\end{align}
In the present scenario, the dominant contribution to $\pi_{ij}^{\mathrm{TT}}$ arises from the fluid kinetic term,
\begin{align}
    \pi_{ij}^{\mathrm{TT}} = (p + \rho)\gamma^2v^{i}v^j,
\end{align}
where $v^i$ is the velocity defined by $v^i=\mathrm{d}x^i/\mathrm{d}\eta$, and $\gamma$ is the Lorentz factor that can be taken as $1$ in the non-relativistic regime. We use $\Tilde{v}_{\vec{k}}^i$ to denote the Fourier transformation of $v^i(\vec{x})$. After bringing it into the UETC, one can obtain a four-point correlation function of $\Tilde{v}_{\vec{k}}^i$. Ignoring the possible non-Gaussianity, it becomes some combination of the two-point correlation function of $\Tilde{v}_{\vec{k}}^i$. After neglecting the rotational component of $v^i(\vec{x})$~\cite{Hindmarsh:2019phv}, the two-point correlation function can be expressed as
\begin{align}
    \langle \Tilde{v}^i_{\vec{k}}(\eta_1)\Tilde{v}^{j*}_{\vec{q}} (\eta_2) \rangle = (2\pi)^3\delta^3(\vec{q}-\vec{k})\hat{q}^i\hat{k}^jB(q, \eta_1, \eta_2).
\end{align}
Here, the assumption of the velocity field being longitudinal is motivated by the results of numerical simulations~\cite{Hindmarsh:2013xza, Hindmarsh:2015qta, Hindmarsh:2017gnf}. After bringing it into Eq.~\eqref{eq:UETC}, one can find
\begin{align}
    \Pi^2(k, \eta_1, \eta_2) = \bar{\omega}^2\int\frac{\mathrm{d}^3q}{(2\pi)^3}B(q, \eta_1, \eta_2)B(\Tilde{q}, \eta_1, \eta_2)\frac{q^2}{\Tilde{q}^2}(1-\mu^2)^2,
\end{align}
where $\bar{\omega} = \bar{\rho}+\bar{p}$ denotes the background's homogeneous value, $\Tilde{q} = |\vec{q}-\vec{k}| $ and $\mu = \hat{q}\cdot\hat{k}$. Then, the next step is to obtain the expression of $B(q, \eta_1, \eta_2)$, which is the task of the next subsection.

\subsection{Fluid velocity field and velocity power spectrum}

On the one hand, the velocity field can be Fourier decomposed as
\begin{align}
    v^i(\eta, \vec{x}) = \frac{1}{2}\int\frac{\mathrm{d}^3q}{(2\pi)^3}[\Tilde{v}^i_{\vec{q}}(\eta)e^{i\vec{q}\cdot\vec{x}} + \Tilde{v}^{i*}_{\vec{q}}(\eta)e^{-i\vec{q}\cdot\vec{x}}],
\end{align}
where $\vec{x}$ and $\vec{q}$ are both comoving quantities. On the other hand, if the sound waves propagate freely after some initial time $\eta_i$, the velocity field also admits the plane wave solutions,
\begin{align}
    v^i(\eta>\eta_i, \vec{x}) = \int\frac{\mathrm{d}^3q}{(2\pi)^3}[v^i_{\vec{q}}e^{-i\omega \eta+i\vec{q}\cdot\vec{x}} + v^{i*}_{\vec{q}}e^{i\omega\eta-i\vec{q}\cdot\vec{x}}],
\end{align}
with $\omega = c_sq$. For longitudinal plane waves, we have $v_{\vec{q}}^i = v_{\vec{q}}\hat{q}^i$. Note that $v^i_{\vec{q}}$ is the amplitude of the plane wave solution, which is different from $\Tilde{v}_{\vec{q}}^i$. In the stationary case\footnote{Stationary means the duration time of this physical process is very long. For the non-stationary case, the reader may refer to Ref.~\cite{RoperPol:2023dzg}.}, one can express $B(q, \eta_1, \eta_2)$ in terms of the two-point function of $v_{\vec{q}}$~\cite{Hindmarsh:2019phv, RoperPol:2023dzg},
\begin{align}\label{eq:defB}
    B(q, \eta_1, \eta_2) = 2P_{v}(q)\cos[\omega(\eta_1-\eta_2)],
\end{align}
where $\langle v_{\vec{q}}v_{\vec{k}}^{*} \rangle \equiv P_{v}(q)(2\pi)^3\delta^3(\vec{k}-\vec{q})$. After defining the energy fluctuation
\begin{align}
    \lambda(x) = \frac{\rho(x)-\bar{\rho}}{\bar{\omega}},
\end{align}
one arrives at the linearized equations of motion for sound waves in Fourier space,
\begin{align}
    \Tilde{\lambda}_{\vec{q}}^{\prime} + iq^j\Tilde{v}^j_{\vec{q}} &= 0,\\
    \Tilde{v}_j^{\prime} + c_siq^j\Tilde{\lambda}_{\vec{q}}&=0.
\end{align}
Combining with their plane wave solutions, one can find that, at some initial time $\eta_i$, the plane wave amplitude is\footnote{The more accurate solution can be found in Ref.~\cite{RoperPol:2023dzg}; however, the final result is almost the same in the stationary case, except for the infrared (IR) behavior.}~\cite{Hindmarsh:2019phv}
\begin{align}
    v_{\vec{q}} = \frac{1}{2}\left( \hat{q}^i\Tilde{v}^i_{\vec{q}}(\eta_i) - c_s\Tilde{\lambda}_{\vec{q}}(\eta_i) \right)e^{i\omega\eta_i}.
\end{align}
Therefore, to derive $P_v(q)$, one should obtain the initial profile of $\Tilde{v}^i_{\vec{q}}(\eta_i)$ and $\Tilde{\lambda}_{\vec{q}}(\eta_i)$. Assuming the single sound shell is spherical and has no interaction with other shells before they collide, one can give rise to the initial profile in coordinate space by superpositions of $n$ sound shells. The velocity profile for the $n$-th sound shell can be written as 
\begin{align}
    v_i^{(n)}(\eta, \vec{x}) = \frac{R_i^{(n)}}{R^{(n)}}u(\eta, R^{(n)}),
\end{align}
where $\vec{R}^{(n)} = \vec{x}-\vec{x}^{(n)}$, and $\vec{x}^{(n)}$ is the center of the $n$-th sound shell and $u(\eta,r)$ is the radial profile of the velocity field. Assuming $\bar{\omega} = \Gamma\bar{\rho}$, one can write the energy fluctuation of the $n$-th sound shell as
\begin{align}
    \lambda^{(n)}(\eta, \vec{x}) = \frac{1}{\Gamma}\delta^{(n)}(\eta, \vec{x}) = \frac{1}{\Gamma}\delta(\eta, R^{(n)}),
\end{align}
where $\delta(\eta, R^{(n)})$ denotes the profile of density contrast. After doing the Fourier transformation, one can obtain
\begin{align}
    \Tilde{v}^{i(n)}_{\vec{q}}(\eta) &= e^{-i\vec{q}\cdot\vec{x}^{(n)}}i\hat{q}^if^{\prime}(\eta,q),\\
    \Tilde{\lambda}^{(n)}_{\vec{q}}(\eta) &= e^{-i\vec{q}\cdot\vec{x}^{(n)}}l(\eta,q),
\end{align}
and functions $f(\eta, q)$, $l(\eta, q)$ are given by
\begin{align}
    f(\eta, q) &= \frac{4\pi}{q}\int_{0}^{\infty}\mathrm{d}r u(\eta, r)\sin (qr),\\
    l(\eta, q) &= \frac{4\pi}{\Gamma q}\int_{0}^{\infty}\mathrm{d}r \delta(\eta, r)r\sin(qr).
\end{align}
After defining a function $A(\eta, r) \equiv (f^{\prime}(\eta, r)+ic_sl(\eta, r))/2$, we have
\begin{align}
    v_{\vec{q}}^{(n)} = ie^{i\omega\eta_i-i\vec{q}\cdot\vec{x}^{(n)}}A(\eta_i, q).
\end{align}
Assuming all the sound shells are propagating freely at roughly the same time $\eta_i$, one can express the two-point correlation function of velocity amplitudes as
\begin{align}
    \langle v_{\vec{q}}^i v_{\vec{k}}^{j*} \rangle = \sum_{m=1}^{N_s}\sum_{n=1}^{N_s} \hat{q}^i\hat{k}^j A(q)A^*(k)\langle e^{-i\vec{q}\cdot\vec{x}^{(m)}+i\vec{k}\cdot\vec{x}^{(n)}} \rangle e^{i(\omega_q -\omega_k )\eta_i},
\end{align}
where $N_s$ denotes the number of sound shells. Then the ensemble average gives rise to
\begin{align}
    \sum_{m=1}^{N_s}\sum_{n=1}^{N_s} \langle e^{-i\vec{q}\cdot\vec{x}^{(m)}+i\vec{k}\cdot\vec{x}^{(n)}} \rangle
    =(2\pi)^3\delta^3(\vec{q}-\vec{k})\frac{N_s}{V_c},
\end{align}
where $V_c$ is the comoving volume of space. Next, we can define the mean comoving separation of the sound shells as $R_{*c} \equiv (V_c/N_s)^{1/3}$. Finally, we arrive at
\begin{align}
    \langle v_{\vec{q}}^i v_{\vec{k}}^{j*} \rangle = \hat{q}^i\hat{q}^j|A(q)|^2(2\pi)^3\delta^3(\vec{q}-\vec{k})\frac{N_s}{V_c}.
\end{align}
Therefore, it yields $P_v(q) = |A(q)|^2/R_{*c}^3$. The dimensionless velocity power spectrum is defined as
\begin{align}
    \mathcal{P}_{v}(q) \equiv \frac{q^3}{2\pi^2}[2P_{v}(q)]
    = \frac{q^3}{\pi^2R_{*c}^3}|A(q)|^2,
\end{align}
where the factor $2$ originates from Eq.~\eqref{eq:defB}. 

\subsection{GW energy spectrum from sound waves}

The Green function in radiation-dominated era\footnote{For the general solution, one can refer to Ref.~\cite{Guo:2020grp}.} is~\cite{Guo:2020grp}
\begin{align}
    G(x, x_0) = \frac{x_0\sin(x-x_0)}{x}.
\end{align}
Then, one can derive the final dimensionless GW energy spectrum,
\begin{align}
\Omega_{\mathrm{GW}}(y,kr_H)= 3\Gamma^2(H_sa_sr_H)\frac{(kr_H)^3}{2\pi^2}
    \Tilde{P}_{\mathrm{GW}}(kr_H)\Upsilon(y),
\end{align}
where $y \equiv a/a_s$ is a time variable, the subscript $s$ denotes the time when the sound waves start sourcing the GWs, roughly the time of sound-shell collisions, $r_H$ is a characteristic length scale (the inital comoving Hubble radius in our scenario), $\Upsilon$ is the usual suppression factor due to Hubble expansion~\cite{Guo:2020grp} that approximates to $1$ at radiation-dominated era and $2/3$ at matter-dominated era in the long-duration limit, and $\Tilde{P}_{\mathrm{GW}}$ is defined as
\begin{align}
    \Tilde{P}_{\mathrm{GW}}(kr_H) = \frac{1}{4\pi c_s kr_H}\left(\frac{1-c_s^2}{c_s^2}\right)^2
    \int_{z_{-}}^{z_{+}}\frac{\mathrm{d}z}{z}\frac{(z-z_{+})^2(z-z_{-})^2}{z_{+}+z_{-}-z}
    \Bar{P}_v(z)\Bar{P}_v(z_{+}+z_{-}-z),
\end{align}
with $z = qr_H$, $z_{\pm} = \frac{kr_H}{2c_s}(1\pm c_s)$, and $\Bar{P}(z) = \frac{\pi^2\mathcal{P}_v(z)}{z^3}$. If we use $R_{*c}$ as the characteristic length scale, we can obtain
\begin{align}
\Omega_{\mathrm{GW}}(y,kr_H)= 3\Gamma^2(H_sa_sR_{*c})\left(\frac{r_H}{R_{*c}}\right)^7\frac{(kr_H)^3}{2\pi^2}\Tilde{P}_{\mathrm{GW}}(kr_H)\Upsilon(y).
\end{align}

This semi-analytical procedure reduces the acoustic GW computations into (i) extracting sound-shell profiles $\delta(r)$ and $u(r)$ (e.g. from 1D GR simulations), (ii) computing $f(k)$, $l(k)$, $A(k)$, and $\mathcal{P}_v(k)$, (iii) evaluating the convolution integral for $\Tilde{P}_{\mathrm{GW}}(k)$, and (iv) applying the multiplicative factors to obtain \red{$\Omega_{\mathrm{GW}}(k)$}. The numerical results and comparisons with 3D lattice evolutions are discussed in Sec.~\ref{sec:results}.

\section{Three-dimensional simulations of acoustic gravitational waves} \label{sec:setup}

In this section, we describe the numerical setup adopted to study the acoustic GW production from an ensemble of curvature peaks. Our GW-generation pipeline proceeds in two stages. The first stage is the gravitational collapse of individual curvature peaks and the consequent excitation of sound waves, which inherently involves strong gravity and hence requires fully GR simulations. The second stage is the collision of the freely propagating sound shells in superpositions and the resulting emission of GWs, which can be treated perturbatively on a fixed FLRW background. An end-to-end, fully GR simulation that resolves both stages is numerically prohibitive. 

Therefore, as a preliminary step, we follow a strategy similar to that in Ref.~\cite{Jinno:2020eqg}: we first (i) perform fully GR, spherically symmetric simulations of the collapse to extract nonperturbative sound-shell profiles and then (ii) embed those spherical profiles into a 3D comoving lattice and evolve the relativistic fluid together with the \red{sourced GWs $h_{ij}$ while neglecting their backreaction on the fluid evolution}, other metric perturbations, and the direct influence of any formed PBHs. This approximation is \red{motivated by the fact that} the energy density carried by sound waves is negligible compared to the total background energy density, and the formed PBHs are sufficiently rare and small that their gravitational fields do not significantly modify the large-scale sound-shell dynamics during the collision phase.

Recent work in the context of cosmological first-order phase transitions also indicates that scalar metric perturbations sourced by sound waves typically contribute negligibly to the GW production compared to the fluid motions themselves~\cite{Giombi:2025tkv}. In subsection~\ref{sec:compare_GW}, we will show that, for our setup, GW emission \red{can receive a dominant contribution from fluid motions rather than from direct gravitational perturbations}. Collisions between sound shells and PBHs could produce additional GWs (analogous to bubble-PBH interactions studied in \cite{Yuwen:2024gcf}); such effects are beyond the scope of the present work.

We assume that the collapse and shell formation occur rapidly compared with the subsequent shell interactions, so that sound shells propagate freely before collisions. Since all curvature peaks are taken to share the same characteristic scale, the sound shells form essentially simultaneously. We extract the sound-shell profile at a chosen time after formation to serve as the initial condition for the 3D lattice simulations. Specifically, we embed $N_s$ non-overlapping copies of this spherical profile at random positions inside a cubic box of comoving side length $L$.

For a comoving volume $V_c=L^3$, the comoving shell number density is
\begin{align}
    n_s=\frac{N_s}{V_c},
\end{align}
or equivalently, the mean comoving separation of shells is
\begin{align}
    R_{*c}=n_s^{-1/3}
    =\left(\frac{V_c}{N_s}\right)^{1/3}
    =\frac{L}{N_s^{1/3}}.
\end{align}
Thus, $N_s$ should not be interpreted as an independent physical control parameter. The quantities $L$ and $N_s$ separately are simulation parameters, while their combination $R_{*c}$ is the physical ensemble parameter controlling the shell abundance once finite-volume effects are controlled.

We implement 3D evolutions using the publicly available lattice code \CosmoLattice~\cite{Figueroa:2021yhd, Figueroa:2020rrl} with an extension that incorporates relativistic hydrodynamics. Following the notation of Refs.~\cite{Figueroa:2023xmq, RoperPol:2025lgc}, the background FLRW metric is written as
\begin{equation}
    \mathrm{d}s^2 = g_{\mu\nu}\mathrm{d}x^\mu \mathrm{d}x^\nu = -a(\eta)^{2\alpha}\mathrm{d}\eta^2+a(\eta)^2\delta_{ij}\mathrm{d}x^i\mathrm{d}x^j,
\end{equation}
where $\eta$ is a time coordinate related to cosmic time $t$ by $\mathrm{d}\eta = a^{-\alpha}\mathrm{d}t$ with constant parameter $\alpha$ (so $\alpha = 0$ corresponds to cosmic time and $\alpha = 1$ to conformal time). In this work, we adopt $\alpha = 1$ so that $\eta$ is conformal time. The energy-momentum tensor of the relativistic fluid is
\begin{equation}
    T^{\mu\nu} = (\rho+p)u^\mu u^\nu + pg^{\mu\nu},
\end{equation}
where $u^\mu$ is the four-velocity with components
\begin{equation}
    u^0 = \gamma/a^{\alpha},\quad u^i = \gamma v^i/a,
\end{equation}
and the comoving three-velocity (with respect to $\eta$) is $v^i = \mathrm{d}x^i/\mathrm{d}\eta$ with corresponding Lorentz factor
\begin{equation}
    \gamma = \frac{1}{\sqrt{1-v^2}} = \frac{1}{\sqrt{1-\delta_{ij}v^iv^j}}.
\end{equation}
This parametrization ensures the normalization $g_{\mu\nu}u^\mu u^\nu = -1$. Using these definitions, the components of $T^{\mu\nu}$ read
\begin{subequations}\label{eq:Tmunu}
    \begin{align}
        T^{00} &= (\rho+p)\gamma^2/a^{2\alpha} - p/a^{2\alpha}, \\
        T^{0i} &= (\rho+p)\gamma^2 v^i/a^{\alpha+1}, \\
        T^{ij} &= (\rho+p)\gamma^2 v^i v^j/a^{2} + p \delta^{ij}/a^{2}.
    \end{align}
\end{subequations}
Energy-momentum conservation, $\nabla_\mu T^{\mu\nu}=0$, yields the fluid evolution equations. For numerical stability and convenience we evolve rescaled components~\cite{Figueroa:2023xmq},
\begin{subequations}
    \begin{align}
        \hat{T}^{00} &= a^{2\alpha+4}T^{00}, \\
        \hat{T}^{0i} &= a^{\alpha+5}T^{0i}, \\
        \hat{T}^{ij} &= a^{6}T^{ij}.
    \end{align}
\end{subequations}
The continuity and momentum equations then become
\begin{subequations}\label{eq:THatEq}
    \begin{align}
        \label{eq:T00} \partial_\eta\hat{T}^{00} + a^{\alpha-1}\partial_i\hat{T}^{0i} + a^4\mathcal{H}(3p - \rho) = 0,\\
        \label{eq:T0i} \partial_\eta\hat{T}^{0i} + a^{\alpha-1}\partial_j\hat{T}^{ij} = 0.
    \end{align}
\end{subequations}
To close the system we express $\hat{T}^{ij}$ in terms of $\hat{T}^{00}$ and $\hat{T}^{0i}$ as
\begin{equation}
    \hat{T}^{ij} = \frac{\hat{T}^{0i}\hat{T}^{0j}}{\hat{T}^{00}+a^4p}+a^4p\delta^{ij}.
\end{equation}
From Eq.~\eqref{eq:Tmunu} one also obtains the algebraic relation
\begin{align}
    \hat{T}^{{00}^2}-\sum_i \hat{T}^{{0i}^2}-a^4(\rho-p)\hat{T}^{00} = a^8\rho p,
\end{align}
which, together with the equation of state $p = \omega\rho$, allows an explicit solution for $a^4p$,
\begin{equation}
    a^4p = \frac{(\omega-1)\hat{T}^{00}+\sqrt{(\omega-1)^2\hat{T}^{{00}^2}-4\omega(\sum_i \hat{T}^{{0i}^2 }- \hat{T}^{{00}^2})}}{2}.
\end{equation}
With $a^4 p$ determined algebraically, Eqs.~\eqref{eq:T00} and~\eqref{eq:T0i} provide closed evolution equations for $\hat{T}^{00}$ and $\hat{T}^{0i}$.

The background FLRW dynamics are evolved self-consistently via the Friedmann equations,
\begin{subequations}
    \begin{align}
        \mathcal{H}^2 &= \left(\frac{a'}{a}\right)^2 = \frac{a^{2\alpha}}{3M_\mathrm{Pl}^2}\langle E\rangle, \\
        \frac{a''}{a} &= \frac{a^{2\alpha}}{6M_\mathrm{Pl}^2}\langle(2\alpha - 1) E - 3P\rangle,
\end{align}
\end{subequations}
where angle brackets denote a spatial volume average, and the effective energy density and pressure are
\begin{subequations}
    \begin{align}
        E &= a^{2\alpha}T^{00} = \hat{T}^{00}/a^4, \\
        P &= a^{2} \sum_i T^{ii} / 3 = \sum_i \hat{T}^{ii} / (3a^4).
    \end{align}
\end{subequations}

The dynamics of GWs then follow from the perturbed Einstein equations,
\begin{equation}
    h_{ij}'' + (3-\alpha)\mathcal{H}h_{ij}' - a^{2\alpha-2}\nabla^2 h_{ij} = 2a^{2\alpha-2}\Pi_{ij}^{\mathrm{TT}},
\end{equation}
where $\Pi_{ij}^{\mathrm{TT}}$ is the TT part of the anisotropic tensor $\Pi_{ij} = T_{ij}-pg_{ij} = T_{ij}-pa^2\delta_{ij}$. The dimensionless GW energy spectrum is defined as in Eq.~\eqref{eq:GW_spctrum}. Since the TT projection is nonlocal and computationally expensive in coordinate space, we evolve an unprojected auxiliary tensor $u_{ij}$ in real space and apply the TT projection in Fourier space when needed. The auxiliary field satisfies
\begin{equation} \label{eq:uij}
    u_{ij}'' + (3-\alpha)\mathcal{H}u_{ij}' - a^{2\alpha-2}\nabla^2 u_{ij} = 2a^{2\alpha-2}\Pi_{ij}^{\mathrm{eff}},
\end{equation}
where the effective anisotropic tensor $\Pi_{ij}^{\mathrm{eff}}$ retains only those components of $\Pi_{ij}$ that have nonzero TT projection; for a relativistic fluid, this reduces to
\begin{equation}
    \Pi_{ij}^{\mathrm{eff}} = a^2(\rho+p)\gamma^2v_iv_j = a^2(\rho+p)\gamma^2v^iv^j = (\hat{T}^{ij}-a^4p\delta^{ij})/a^2.
\end{equation}
In \CosmoLattice, the second-order Eq.~\eqref{eq:uij} is cast as a first-order system by introducing the conjugate momentum tensor $\pi_{ij} = a^{3-\alpha} u_{ij}'$, yielding
\begin{subequations}
    \begin{align}
        u_{ij}' &= a^{\alpha-3}\pi_{ij}, \\
        \pi_{ij}' &=  a^{1+\alpha}\nabla^2 u_{ij} + 2a^{1+\alpha}\Pi_{ij}^{\mathrm{eff}}.
    \end{align}
\end{subequations}

For the simulations presented here, we adopt radiation domination ($\omega = 1/3$). To implement the system numerically, we rescale physical variables to dimensionless program variables,
\begin{equation}
    \tilde{\eta} = \omega_* \eta, \quad \tilde{x} = \omega_* x, \quad \tilde{T}_{\mu\nu} = \frac{\hat{T}_{\mu\nu}}{f_*^2\omega_*^2}, \quad \tilde{u}_{ij} = \left(\frac{M_\mathrm{Pl}}{f_*}\right)^2 u_{ij},
\end{equation}
where $\omega_*$ and $f_*$ are free scaling parameters. In this work, we set $\omega_* = 1/(a_1r_H)$ and $f_* = 1$, where $a_1$ is the scale factor at the time when the 1D sound-shell profile is extracted. The coupled fluid-GW system is evolved on a 3D comoving lattice with periodic boundary conditions. The hydrodynamic equations are integrated with a fourth-order Runge-Kutta scheme, the GW equations with a second-order leapfrog integrator, and spatial derivatives are discretized using fourth-order central finite differences.

\begin{figure}[htbp]
    \centering
    \includegraphics[width=\linewidth]{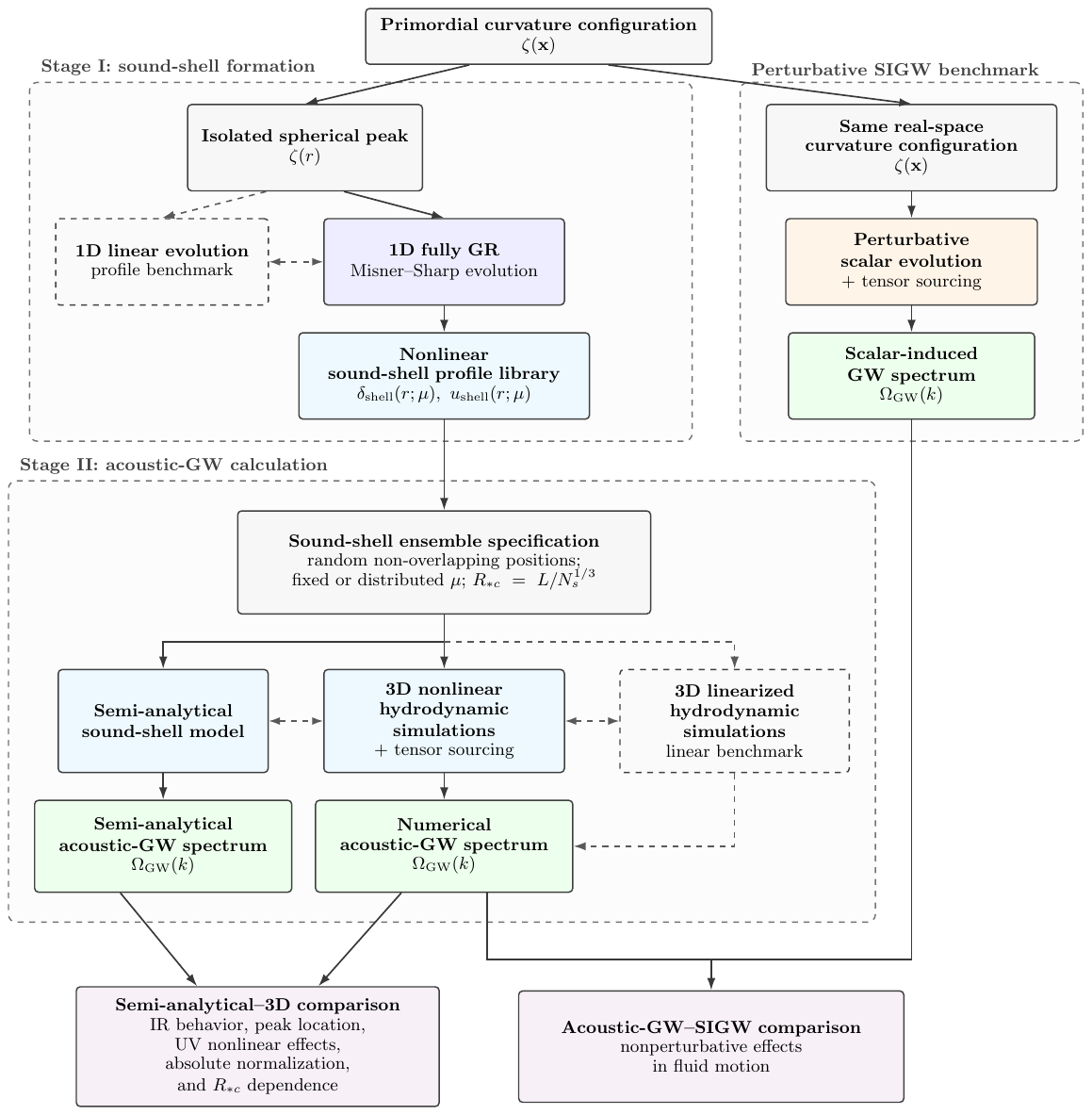}
    \caption{\red{Detailed workflow of the calculation and its cross-checks. Stage I constructs the nonlinear sound-shell profile library as a function of $\mu$ and compares the profiles with linear evolution. Stage II uses the selected profiles in the semi-analytical model and 3D hydrodynamic simulations. The separate right-hand branch gives the perturbative SIGW benchmark. Dashed benchmark boxes and one-way dashed arrows identify the linear benchmarks, while dashed double-headed arrows denote direct comparisons or cross-checks.}}
    \label{fig:workflow_detailed}
\end{figure}

\red{Figure~\ref{fig:workflow_detailed} summarizes the intermediate steps and cross-checks developed in Secs.~\ref{sec:SW}--\ref{sec:setup}, including the comparison between fully GR and linear sound-shell evolution, the semi-analytical and 3D acoustic-GW calculations, and the linearized-hydrodynamics and SIGW benchmarks.}

\section{Numerical results} \label{sec:results}

In this section, we present 3D lattice numerical simulation results for the GW energy spectra sourced by sound-shell collisions. Unless stated otherwise, our simulations use a cubic grid of size $N^3 = 512^3$, a dimensionless space interval $\mathrm{d}\tilde{x} = \tilde{L} / N$, and a time step $\mathrm{d}\tilde{\eta} = 0.2\mathrm{d}\tilde{x}$. The initial scale factor is set to $a(\tilde{\eta}_i) = 1$ and the initial conformal Hubble parameter $\tilde{\mathcal{H}}_i$ is taken from the 1D simulation at the time when the sound-shell profile is extracted; the initial conformal time is therefore $\tilde{\eta}_i = 1/\tilde{\mathcal{H}}_i$. Each simulation is evolved until $\tilde{\eta}_{\mathrm{max}} = \tilde{L}$, which is sufficient for the GW spectrum to converge while keeping the comoving Hubble radius smaller than the box size. The comoving box length $\tilde{L}$ and the number of embedded sound shells $N_s$ are specified for each case below.

\begin{figure}[htbp]
    \centering
    \begin{subfigure}{0.24\linewidth}
        \includegraphics[height=\linewidth]{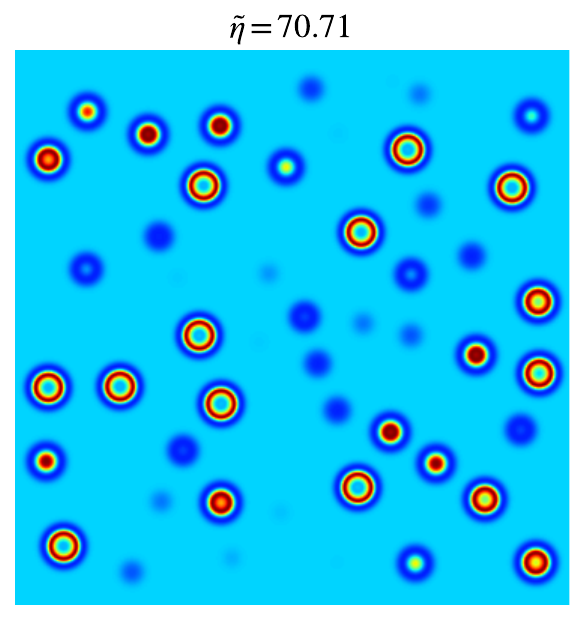}
        \caption{}
    \end{subfigure}
    \begin{subfigure}{0.24\linewidth}
        \includegraphics[height=\linewidth]{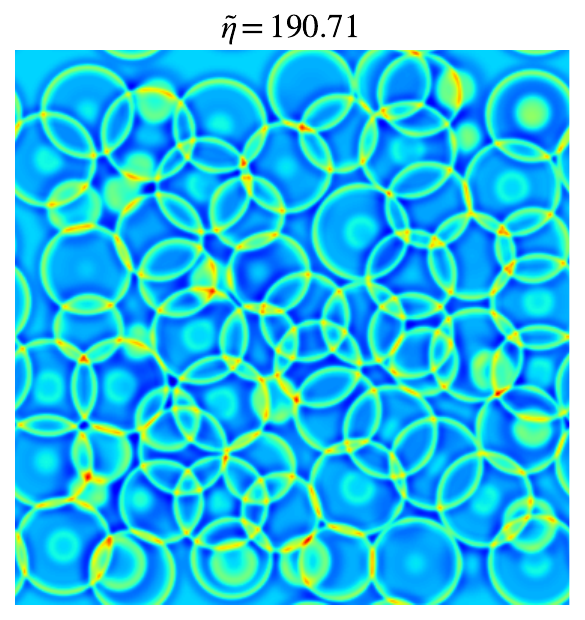}
        \caption{}
    \end{subfigure}
    \begin{subfigure}{0.24\linewidth}
        \includegraphics[height=\linewidth]{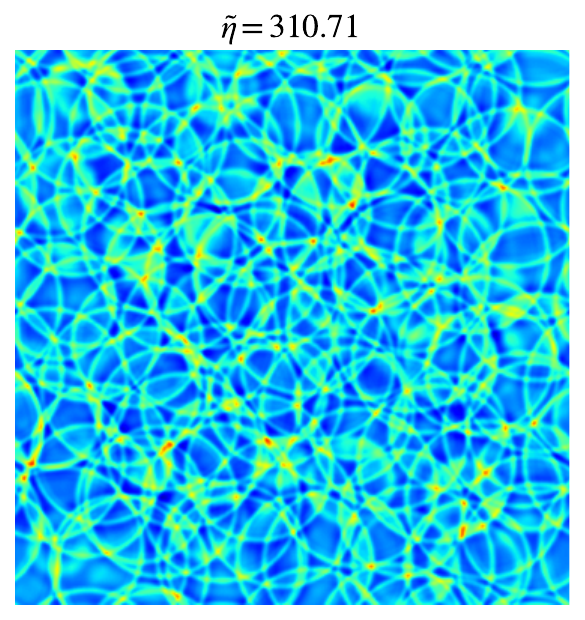}
        \caption{}
    \end{subfigure}
    \begin{subfigure}{0.24\linewidth}
        \includegraphics[height=\linewidth]{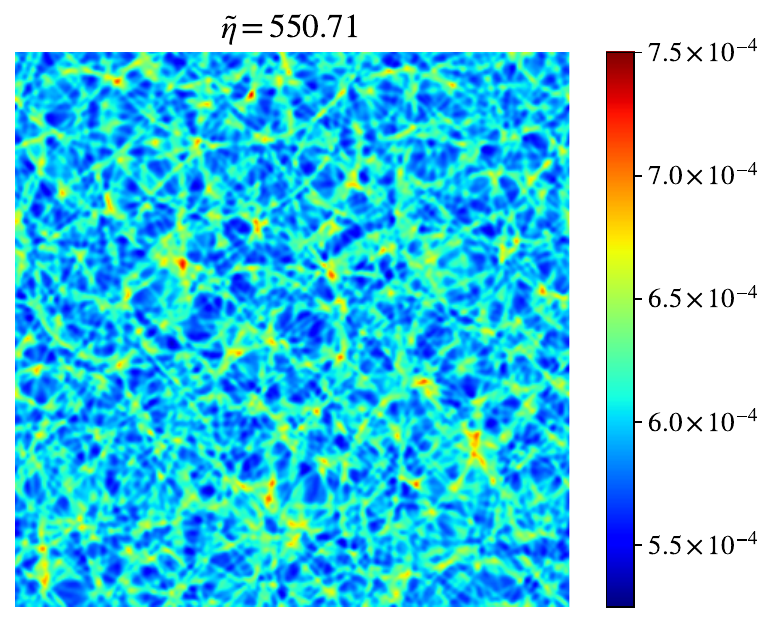}
        \caption{}
    \end{subfigure}
    \caption{Slices of $\tilde{T}^{00}$ from a representative simulation. The initial condition is constructed by embedding the 1D sound-shell profile with $\mu = 0.4$ at $t_i = 50t_m$; the box length is $\tilde{L} = 1200$, and the number of sound shells is $N_s = 400$. The four snapshots correspond to $\tilde{\eta} = 70.71$ (a), $\tilde{\eta} = 190.71$ (b), $\tilde{\eta} = 310.71$ (c), and $\tilde{\eta} = 550.71$ (d).}
    \label{fig:slices}
\end{figure}

Fig.~\ref{fig:slices} displays representative slices of $\tilde{T}^{00}$ from a typical run. In this example, the 3D initial condition is constructed by embedding the 1D sound-shell profile with $\mu = 0.4$ at $t_i = 50t_m$ into a box of length $\tilde{L} = 1200$ with $N_s = 400$ shells. The corresponding initial conformal time is $\tilde{\eta}_i = 50\sqrt{2} \approx 70.71$. The four panels correspond to $\tilde{\eta} = 70.71$ (a), $\tilde{\eta} = 190.71$ (b), $\tilde{\eta} = 310.71$ (c), and $\tilde{\eta} = 550.71$ (d).

\subsection{Shape of the acoustic GW energy spectrum}

We first consider a simplified scenario in which every curvature peak has the same amplitude $\mu$, enabling direct comparisons with the semi-analytical results from the sound shell model~\cite{Zeng:2025law}. We choose four representative amplitudes: subcritical ($\mu = 0.4$), negative ($\mu = -0.4$), near-critical ($\mu = 0.8$), and supercritical ($\mu = 0.9$).

\begin{figure}[htbp]
    \centering
    \includegraphics[width=0.45\linewidth]{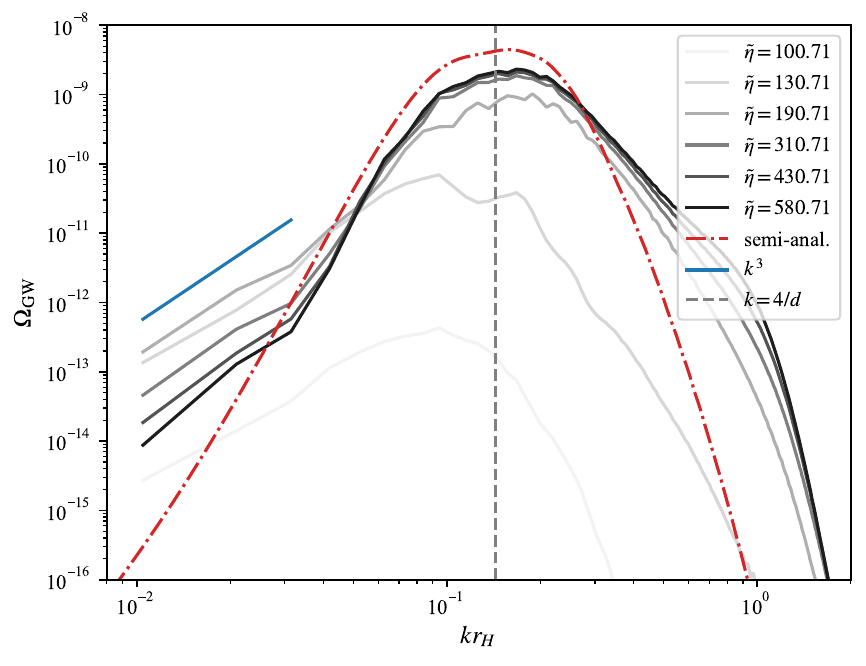}
    \qquad
    \includegraphics[width=0.45\linewidth]{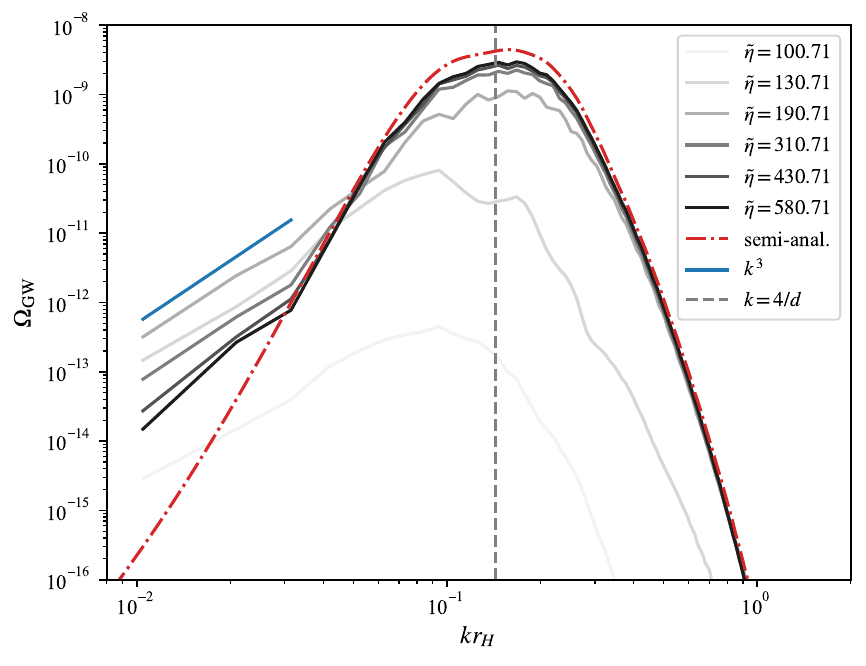}
    \\
    \includegraphics[width=0.45\linewidth]{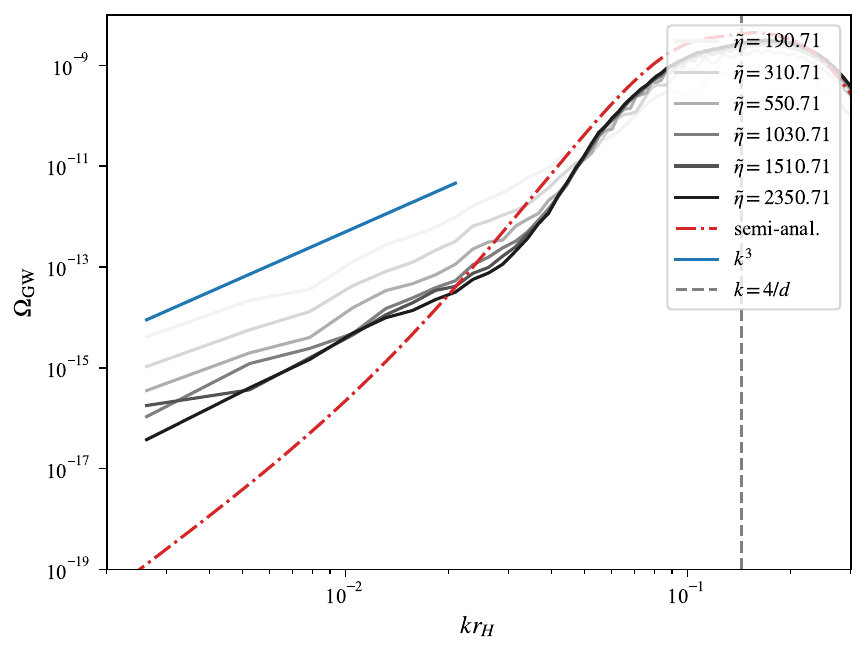}
    \qquad
    \includegraphics[width=0.45\linewidth]{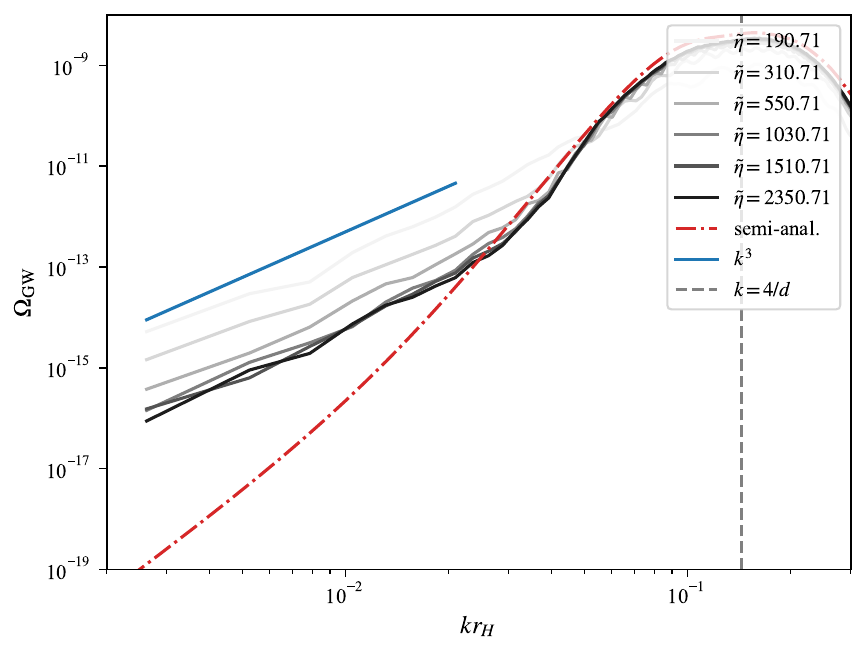}
    \caption{\red{GW energy spectra at several time slices for the subcritical case ($\mu=0.4$). Top row: $\tilde{L}=600$, $N_s=50$. Bottom row: $\tilde{L}=2400$, $N_s=3200$. Left panels use the nonlinear hydrodynamic equations, while right panels use the linearized hydrodynamics. The initial condition is constructed from the 1D sound-shell profile at $t_i=50t_m$. The red dot-dashed lines show the semi-analytical results from the sound shell model with their absolute normalization; no rescaling to the numerical spectra has been applied. The dashed vertical line marks $k=4/d$, where $d$ is the comoving shell thickness. The blue reference line indicates the causal $k^3$ scaling in the IR regime.}}
    \label{fig:subcritical}
\end{figure}

\begin{figure}[htbp]
    \centering
    \includegraphics[width=0.32\linewidth]{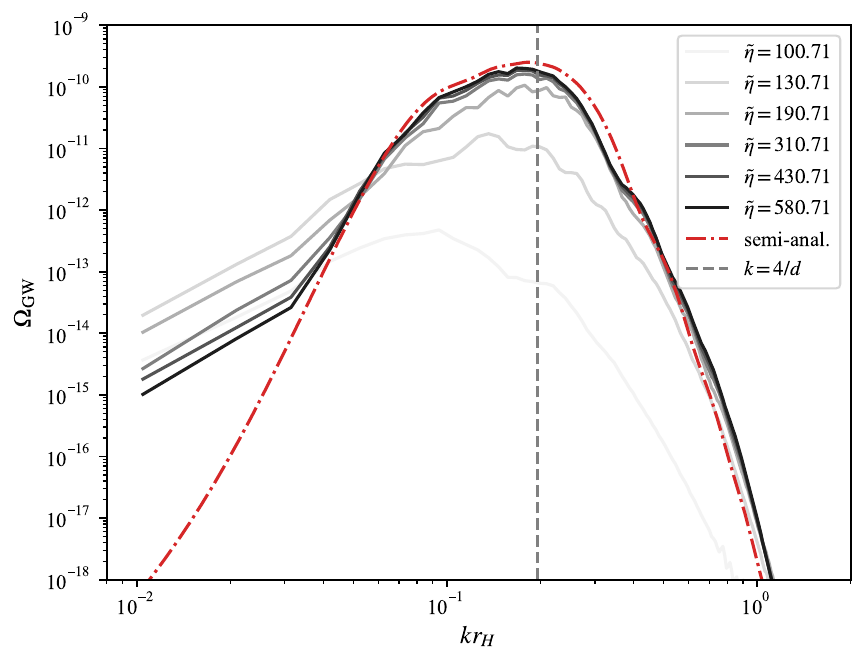}
    \includegraphics[width=0.32\linewidth]{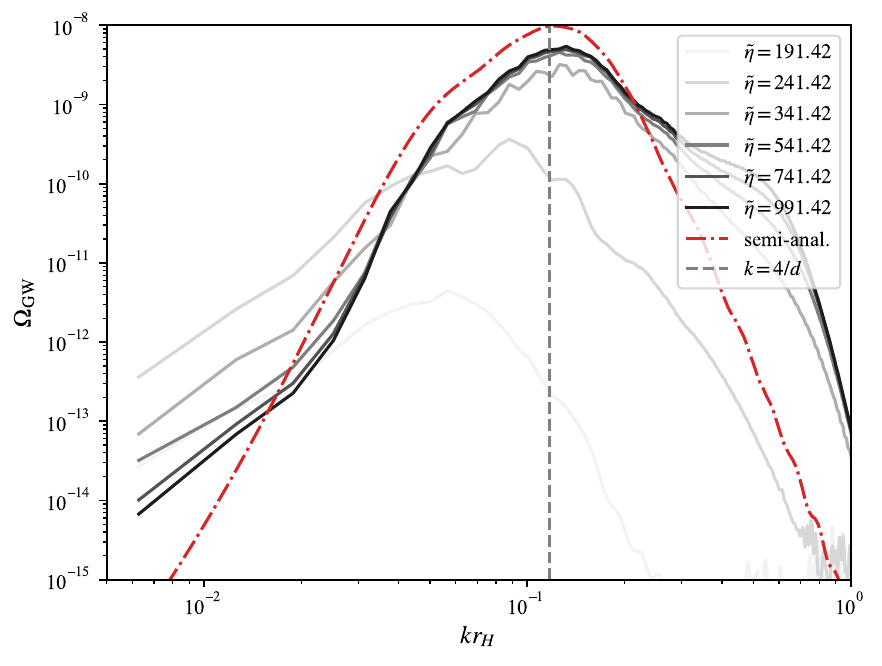}
    \includegraphics[width=0.32\linewidth]{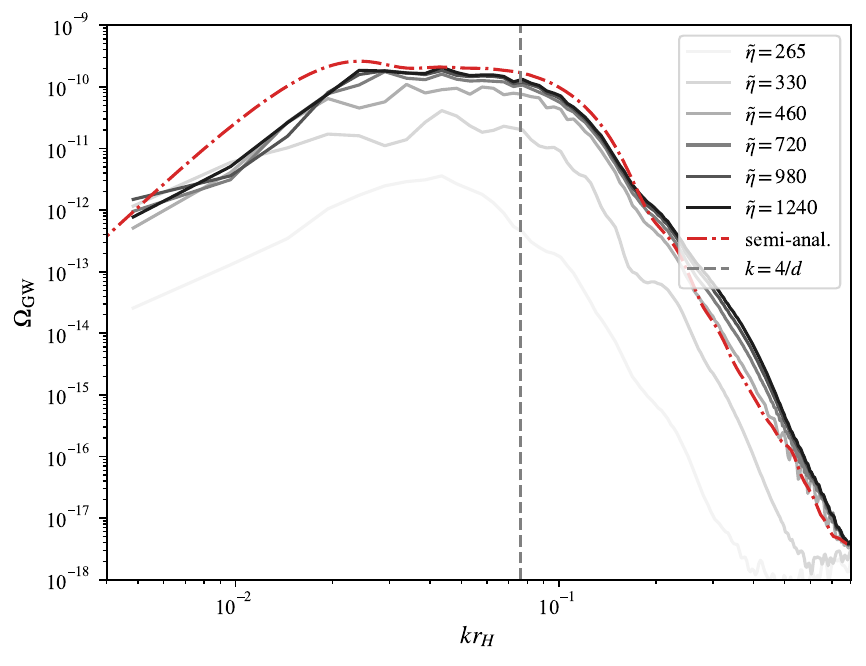}
    \\
    \includegraphics[width=0.32\linewidth]{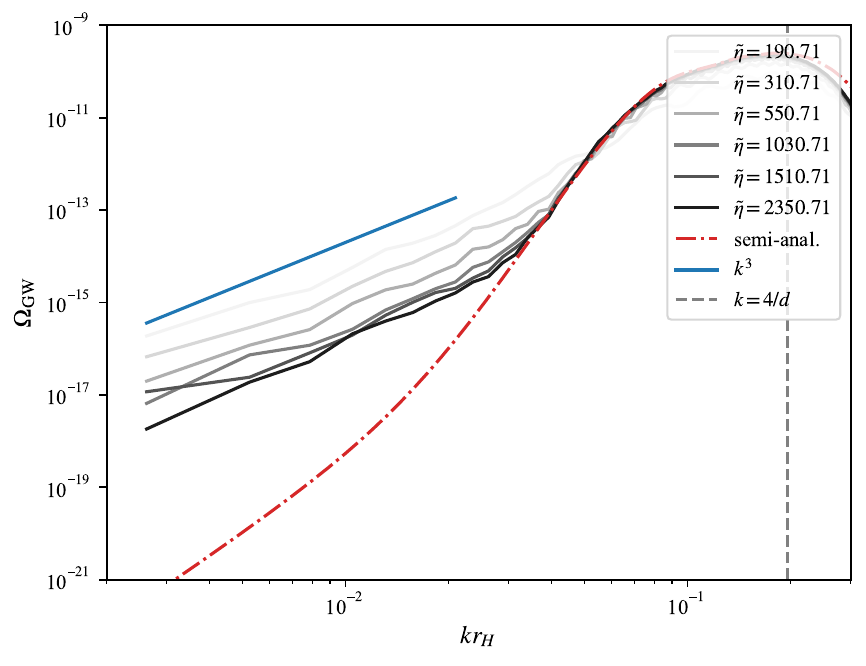}
    \includegraphics[width=0.32\linewidth]{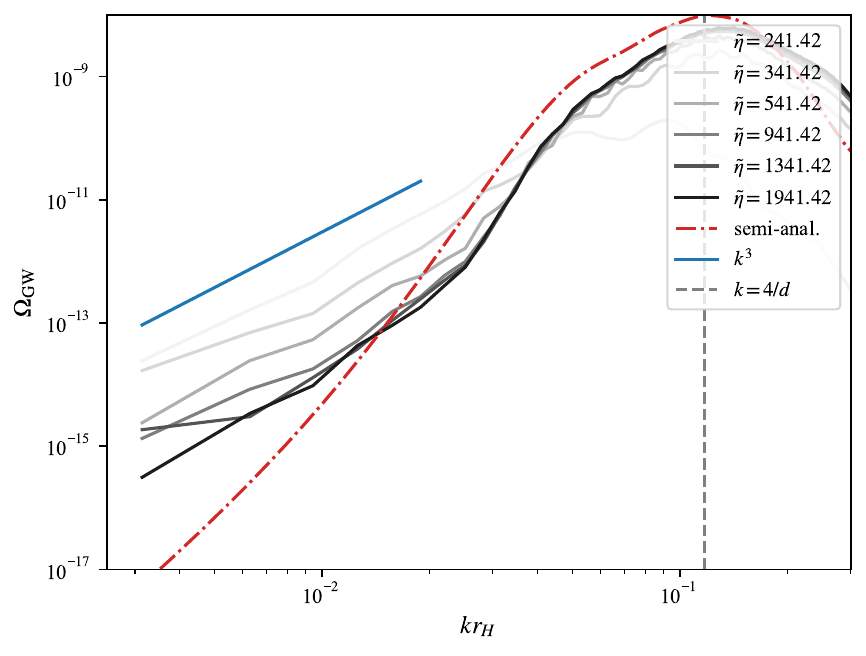}
    \includegraphics[width=0.32\linewidth]{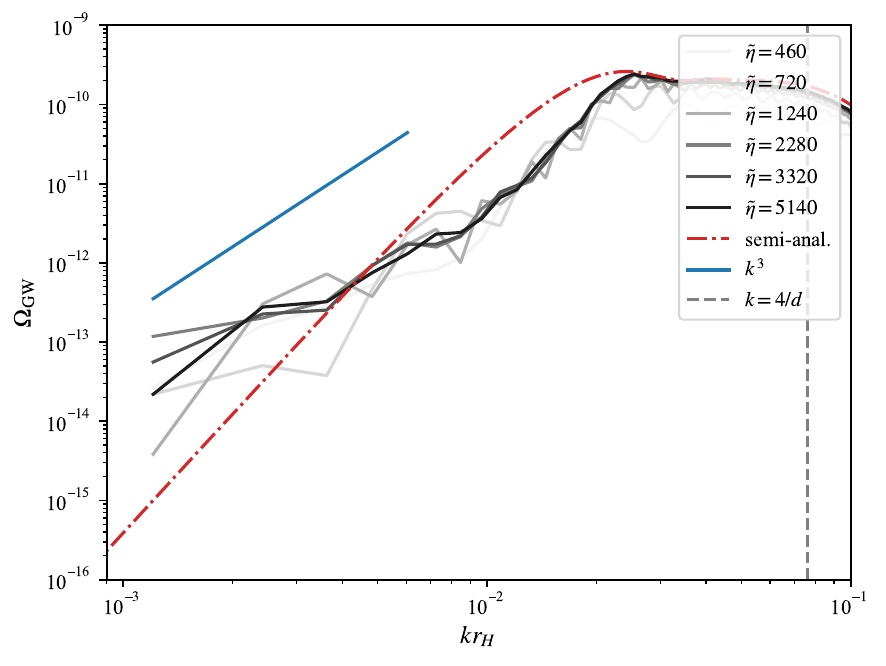}
    \caption{\red{GW energy spectra at several time slices for negative-amplitude ($\mu=-0.4$, left), near-critical ($\mu=0.8$, middle), and supercritical ($\mu=0.9$, right) perturbations. The red dot-dashed lines show the semi-analytical results from the sound shell model with their absolute normalization; no peak matching or additional rescaling has been applied. Top row: representative small-box runs with $\tilde L=600$, $1000$, and $1300$, respectively, and $N_s=50$. Bottom row: larger-box, higher-$N_s$ simulations used to confirm the IR scaling ($\tilde L=2400$, $N_s=3200$ for negative-amplitude; $\tilde L=2000$, $N_s=400$ for near-critical; $\tilde L=5200$, $N_s=3200$ for supercritical).}}
    \label{fig:others}
\end{figure}

We begin with the subcritical case, which produces no PBHs and thus permits direct use of the 1D sound-shell profile as the 3D initial condition. A representative run uses the 1D profile at $t_i = 50t_m$, a box length $\tilde{L} = 600$, and $N_s = 50$ sound shells. The GW energy spectra at several time slices are shown in the top-left panel of Fig.~\ref{fig:subcritical}, together with the semi-analytical result (red dot-dashed) from the sound shell model. \red{All semi-analytical spectra in Figs.~\ref{fig:subcritical} and~\ref{fig:others} are shown with their absolute normalization; no rescaling to the numerical spectra has been applied. We therefore distinguish agreement in spectral shape and peak position from agreement in absolute normalization. The characteristic peak position continues to track the shell thickness through $k\simeq4/d$, marked by the vertical dashed line. Systematic differences appear in both the infrared (IR) and ultraviolet (UV) regimes.}

\red{For a given peak amplitude $\mu$, we increase $\tilde L$ and $N_s$ together to keep $R_{*c}$ fixed. For the subcritical and negative-amplitude setups, $(\tilde L,N_s)=(600,50)$ and $(2400,3200)$ both give $R_{*c}\simeq162.9r_H$. For the near-critical case, $(1000,50)$ and $(2000,400)$ both give $R_{*c}\simeq271.4r_H$, while the supercritical case uses $(1300,50)$ and $(5200,3200)$, both giving $R_{*c}\simeq352.9r_H$. These paired simulations therefore leave the comoving shell number density unchanged and provide finite-volume tests at fixed physical shell abundance. Increasing the box size extends the accessible range toward smaller wavenumbers and is used in particular to verify that the causal $k^3$ infrared behavior is not a small-box artifact.}

In the IR regime, the numerical spectra exhibit a causal tail proportional to $k^3$, rather than the steeper $k^5$ scaling from the semi-analytical estimate. We attribute this discrepancy to the finite lifetime of the GW source, as the semi-analytical derivation already assumes a long-duration limit. Refs.~\cite{RoperPol:2023dzg,Cai:2019cdl} show that including a finite source duration modifies the IR scaling to a causal $k^3$ behavior, which is consistent with our numerical findings. We defer a detailed semi-analytical investigation of finite-duration effects to future work. We confirm the robustness of the $k^3$ IR tail by repeating the simulation with a larger box ($\tilde{L} = 2400$) and more shells ($N_s = 3200$) in the bottom-left panel of Fig.~\ref{fig:subcritical} that shows the same $k^3$ behavior.

In the UV region, our GW spectra exceed the semi-analytical prediction and show a pronounced bump immediately beyond the peak. We attribute this enhancement to nonlinear hydrodynamic interactions that are absent from the linear superposition assumed in the sound shell model. To test this interpretation, we perform companion simulations using the linearized hydrodynamic equations by trimming off the nonlinear interactions from Eqs.~\eqref{eq:THatEq},
\begin{subequations}
    \begin{align}
        3\delta' + 4\partial_i v^i &= 0, \\
        4{v^i}' + \partial_i\delta &= 0,
    \end{align}
\end{subequations}
with the same initial conditions and parameters. \red{The right panels of Fig.~\ref{fig:subcritical} show that, once nonlinear hydrodynamic interactions are removed, the numerical spectrum follows the semi-analytical spectral shape more closely, particularly in the UV, while a residual absolute-normalization difference remains. This comparison indicates that nonlinear hydrodynamics modifies both the UV shape and, to some extent, the peak amplitude, but does not account for the entire normalization difference. The IR behavior is controlled by causal effects and remains robust to nonlinearities.}

The upper-left panel of Fig.~\ref{fig:others} shows the negative-amplitude case ($\mu=-0.4$) for which we use $t_i = 50t_m$, $\tilde{L} = 600$ and $N_s = 50$. For near-critical and supercritical perturbations, a PBH forms at the center while a sound shell propagates outward. Since the high-density, gravitationally bound region surrounding the PBH does not participate in outward propagation, we excise the central portion of the 1D profile before embedding it into the 3D lattice. We have verified that modest variations of the cutoff radius produce only marginal changes in the GW spectra; in the results below, the excision radius is chosen at the first zero of the density contrast, and the excised region is replaced by the homogeneous background. Initial profiles are taken at $t_i = 200t_m$ for the near-critical case, and at $t_i = 400t_m$ for the supercritical case. These runs use $\tilde{L} = 1000$ and $\tilde{L} = 1300$, respectively, with $N_s = 50$ shells. The middle and right panel of the top row in Fig.~\ref{fig:others} displays these spectra together with semi-analytical predictions. The bottom row presents corresponding larger-box, higher-$N_s$ runs used to verify the robustness of IR behavior ($\tilde{L} = 2400$, $N_s = 3200$ for negative-amplitude; $\tilde{L} = 2000$, $N_s = 400$ for near-critical; $\tilde{L} = 5200$, $N_s = 3200$ for supercritical). \red{In all cases, the semi-analytical spectra are shown with their absolute normalization, without peak matching or additional rescaling.}

\red{The different baseline box sizes used for different values of $\mu$ reflect the different spatial extents of the sound-shell profiles used as the 3D initial conditions. The subcritical and negative-amplitude profiles are extracted at $t_i=50t_m$, the near-critical profile at $t_i=200t_m$, and the supercritical profile at $t_i=400t_m$. Their radial extents and shell thicknesses also depend on $\mu$. Since the 3D initial conditions are constructed by embedding $N_s$ non-overlapping copies of these profiles, the larger profiles require correspondingly larger simulation boxes. Consequently, the representative runs in Figs.~\ref{fig:subcritical} and~\ref{fig:others} are primarily used to compare the spectral morphology and to test the infrared behavior for each $\mu$, rather than to isolate the absolute $\mu$-dependence of the GW amplitude at fixed $R_{*c}$. Direct comparisons of the amplitude dependence on $\mu$ are instead performed in Sec.~\ref{sec:compare_GW} using matched real-space configurations.}

To quantify the absolute-normalization difference, we define
\begin{align}
    \mathcal R_X\equiv
    \frac{\Omega_{\rm GW,X}^{\rm peak}}
         {\Omega_{\rm GW,semi}^{\rm peak}},
    \qquad X={\rm NL},{\rm lin},
\end{align}
where ``NL'' and ``lin'' denote the nonlinear and linearized 3D hydrodynamic simulations, respectively, and ``semi'' denotes the semi-analytical sound-shell result without rescaling. Each peak is determined independently from the corresponding spectrum. The measured ratios are listed in Table~\ref{tab:peak_ratios}.

\begin{table}[htbp]
    \centering
    \caption{\red{Ratios of the numerical to semi-analytical GW peak amplitudes. Each numerical and semi-analytical peak is determined independently. The paired small- and large-box simulations in each row have the same mean comoving shell separation $R_{*c}$.}}
    \label{tab:peak_ratios}
    \begin{tabular}{c c c c c}
        \hline\hline
        Case & Small-box $(\tilde L,N_s)$ & $\mathcal R^{\rm small}$ & Large-box $(\tilde L,N_s)$ & $\mathcal R^{\rm large}$ \\
        \hline
        $\mu=0.4$, NL  & $(600,50)$  & $0.523$ & $(2400,3200)$ & $0.707$ \\
        $\mu=0.4$, lin & $(600,50)$  & $0.664$ & $(2400,3200)$ & $0.776$ \\
        $\mu=-0.4$, NL & $(600,50)$  & $0.806$ & $(2400,3200)$ & $0.920$ \\
        $\mu=0.8$, NL  & $(1000,50)$ & $0.554$ & $(2000,400)$  & $0.635$ \\
        $\mu=0.9$, NL  & $(1300,50)$ & $0.796$ & $(5200,3200)$ & $0.927$ \\
        \hline\hline
    \end{tabular}
\end{table}

\red{The absolute difference is not described by a universal factor: it depends on the physical case and on the simulation volume. For the subcritical case in Fig.~\ref{fig:subcritical}, the nonlinear and linearized numerical peaks lie below the semi-analytical peak by $47.7\%$ and $33.6\%$, respectively, in the small box, and by $29.3\%$ and $22.4\%$ in the large box. The large-box ratio is systematically closer to unity than the corresponding small-box ratio when $R_{*c}$ is held fixed. This trend is consistent with the mismatch in the small boxes being partly caused by finite-volume and finite-realization effects. A larger volume contains more shells, improving the statistical sampling of the random shell ensemble, and reduces sensitivity to the infrared cutoff and periodicity of the box. These results do not establish an infinite-volume extrapolation or show that the ratio must converge exactly to unity. Residual differences may reflect the ensemble/statistical treatment of shell positions versus a finite realization, the long-duration approximation versus the finite source duration, the characteristic source time in the semi-analytical estimate, and, in the nonlinear calculation, hydrodynamic interactions absent from the linear sound-shell model. We therefore do not attribute the residual normalization difference to any single effect without isolating it quantitatively.}

Recall that the negative-amplitude case reverses the radial ordering of overdense and underdense shells relative to the subcritical case, but its overall spectral shape remains similar. The near-critical spectra also resemble the subcritical result, due to similar sound-shell structure. The numerical and semi-analytical spectra agree well near the peak but diverge in the IR and UV regions. In the supercritical runs, the profile contains only an underdense shell, producing a broader, plateau-like spectral peak whose characteristic frequency lies near $k = 4/d$, where $d$ is the comoving thickness of the underdense shell for the supercritical case. Notably, for the negative-amplitude and supercritical runs, the enhancement of the numerical UV regime is not obvious, and the spectra are consistent with the semi-analytical prediction. This consistency follows from the smaller shell amplitudes in those cases, which suppress nonlinear hydrodynamic effects. Across all cases, the IR regime consistently exhibits the causal $k^3$ tail.

\subsection{Effects of the mean comoving separations of sound shells}

In Sec.~\ref{sec:semi}, we have estimated the GW energy spectrum as
\begin{align}
\Omega_{\mathrm{GW}}(y,kr_H)= 3\Gamma^2(H_sa_sR_{*c})\left(\frac{r_H}{R_{*c}}\right)^7\frac{(kr_H)^3}{2\pi^2}\Tilde{P}_{\mathrm{GW}}(kr_H)\Upsilon(y).
\end{align}
The factor $\frac{(kr_H)^3}{2\pi^2}\Tilde{P}_{\mathrm{GW}}(kr_H)$ determines the spectra shape, and its amplitude is largely insensitive to the collision time $t_c$ (see Fig. 3 of Ref.~\cite{Zeng:2025law}). The combination $H_sa_sR_{*c}$ is also approximately independent of $t_c$, since $H_sa_s \sim 1/\sqrt{2r_Ht_c}$ while the mean comoving separation grows like $R_{*c} \sim 2\Delta r_s \sim 2c_s\Delta\eta \sim 2c_s\sqrt{2r_Ht_c}$ with $\Delta r_s$ the comoving distance traveled by shells before collision. Therefore, the overall amplitude is expected to be effectively independent of the collision time and scales as $R_{*c}^{-7}$.

\begin{figure}[htbp]
    \centering
    \includegraphics[width=0.9\linewidth]{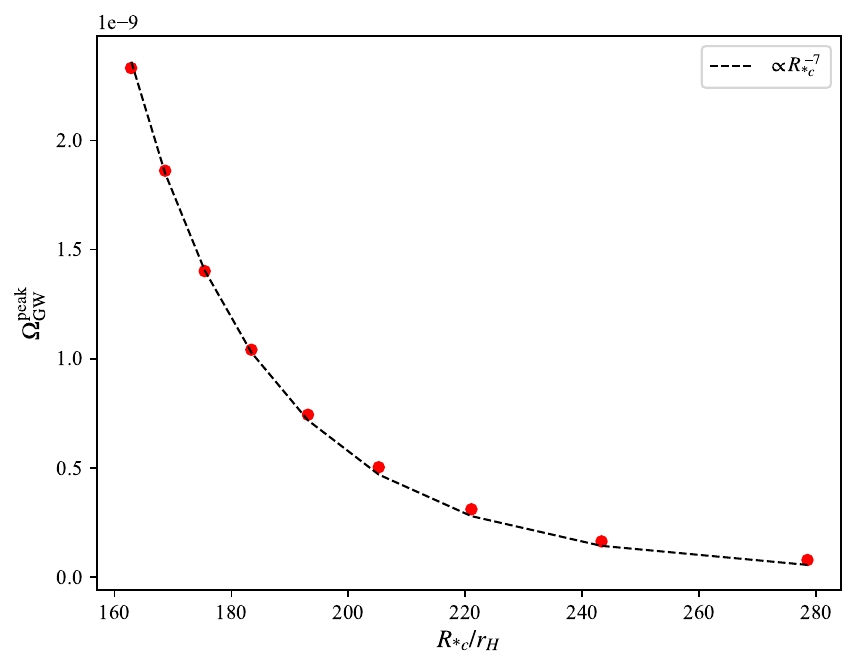}
    \caption{Peak amplitudes of the GW energy spectra as a function of the mean comoving separation $R_{*c}$. Initial conditions use the 1D profile with $\mu = 0.4$ at $t_i = 50t_m$. The box length is $\tilde{L} = 600$; $N_s$ is varied over ${10,15,20,25,30,35,40,45,50}$ to change $R_{*c}$. The dashed line is a power-law fit $\propto R_{*c}^{-7}$, showing agreement with the analytical expectation.}
    \label{fig:amplitude}
\end{figure}

We test this prediction numerically by varying $R_{*c}$ in the 3D simulations. \red{Unlike the finite-volume tests above, here we deliberately vary the physical shell abundance.} Concretely, we fix the box length to $\tilde{L} = 600$, and construct 3D initial conditions from the 1D sound-shell profile with $\mu = 0.4$ extracted at $t_i = 50t_m$, then we vary the number of sound shells $N_s$ to adjust $R_{*c}$. Each configuration is evolved until $\tilde{\eta}_{\mathrm{max}} = \tilde{L}$, and finally, we measure the GW peak amplitude. The measured peak amplitudes are plotted against $R_{*c}$ in Fig.~\ref{fig:amplitude}. The results follow a power law consistent with the $R_{*c}^{-7}$ scaling (the dashed line is the best fit), confirming our analytical expectation.

Because of this steep dependence, the integrated GW amplitude is extremely sensitive to the mean separation of sound shells: even modest individual shell amplitudes can yield a large collective GW signal when shells are sufficiently closely spaced.

\subsection{Effects of amplitude distributions of perturbations}\label{sec:amplitude_distribution}

\begin{figure}[htbp]
    \centering
    \includegraphics[width=0.45\linewidth]{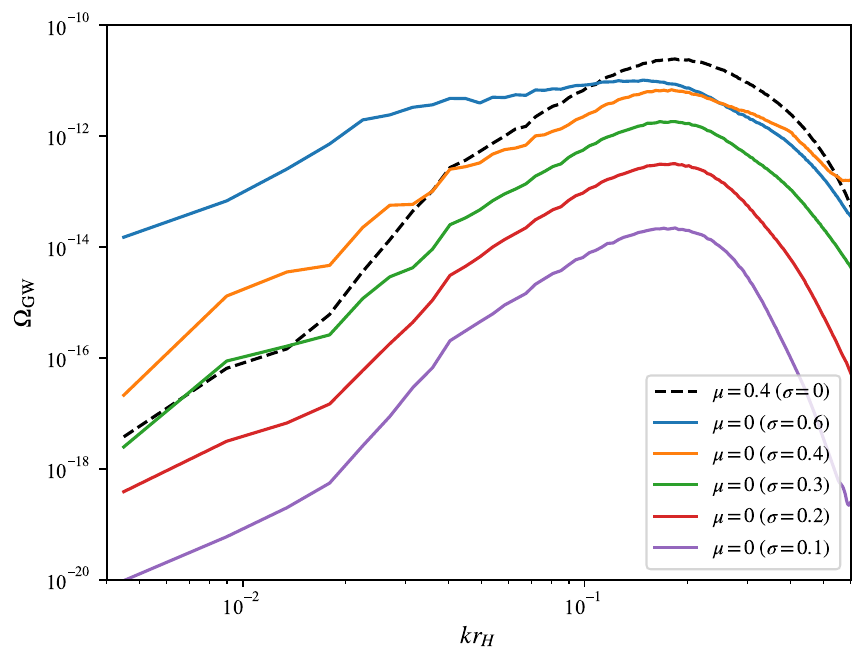}
    \qquad
    \includegraphics[width=0.45\linewidth]{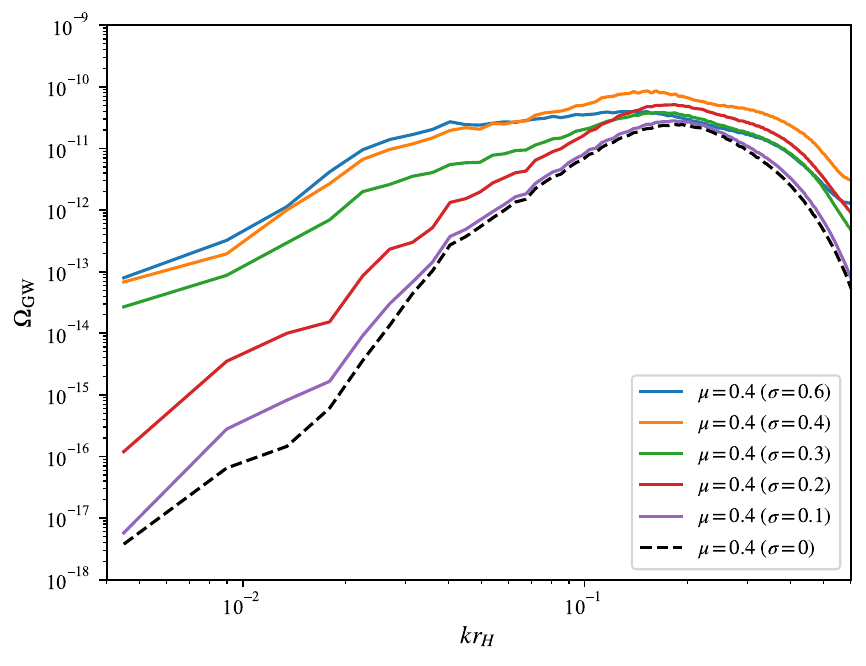}
    \caption{Final GW energy spectra for curvature perturbations with amplitudes sampled from a Gaussian distribution of standard deviation $\sigma = 0.1$, $0.2$, $0.3$, $0.4$, and $0.6$. Left panel: zero-mean distribution. Right panel: mean of $\mu = 0.4$. Initial conditions are constructed from the 1D sound-shell profile at $t_i = 200t_m$. The box length is $\tilde{L} = 1400$, and the number of sound shells is $N_s = 100$. The black dashed line shows the spectrum from the single-amplitude case $\mu = 0.4$ ($\sigma = 0$) with other parameters unchanged for comparison.}
    \label{fig:sigma}
\end{figure}

So far, we have considered collisions of sound shells produced by perturbations of a single amplitude $\mu$. In reality, however, the curvature perturbation amplitudes follow a distribution, and the resulting sound-shell profiles depend nonlinearly on $\mu$: perturbations closer to the critical threshold produce larger-amplitude shells, whereas supercritical perturbations generate only underdense shells. Therefore, a realistic treatment should consider an amplitude distribution for the curvature perturbations.

\red{Here we relax only the identical-amplitude assumption, while all curvature peaks are still taken to share the common characteristic scale $r_m$.} As a first step toward a more realistic treatment, we model $\mu$ with a zero-mean Gaussian distribution of standard deviation $\sigma$. \red{We use rejection sampling, discarding draws with $|\mu| > 1.28$ until $N_s$ accepted amplitudes are obtained.} This truncation restricts the sample to type-I perturbations\footnote{For recent discussions of type-I and type-II perturbations, see Refs.~\cite{Uehara:2024yyp, Harada:2024trx, Escriva:2025eqc}.}. Type-II perturbations are rare for small $\sigma$ and are neglected here for simplicity. For each accepted $\mu$, we use the corresponding 1D sound-shell profile at $t_i = 200t_m$ as the 3D initial condition. The runs reported below use $\tilde{L} = 1400$ and $N_s = 100$, and then we vary $\sigma$ and compute the resulting GW spectra.

The left panel of Fig.~\ref{fig:sigma} presents the final GW energy spectra for \red{$\sigma = 0.1$ (purple), $0.2$ (red), $0.3$ (green), $0.4$ (orange), and $0.6$ (blue)}. For reference, we also plot the spectrum from the single-amplitude case $\mu = 0.4$ with other parameters unchanged (black dashed line). The IR behavior is universal and unaffected by the amplitude distribution, consistently exhibiting the causal $k^3$ scaling. The peak structure and UV behavior, however, vary with $\sigma$. For small $\sigma$, most samples are subcritical, and the spectrum closely resembles the single-amplitude subcritical case. As $\sigma$ increases, the fraction of relatively large-amplitude shells grows, the overall GW amplitude increases, and the UV tail becomes slightly shallower (e.g. \red{$\sigma = 0.4$, orange curve}). These trends are expected: larger $\sigma$ increases the contribution of larger shells, which enhances nonlinear hydrodynamic interactions that both amplify the UV tail and boost the integrated GW power. When $\sigma$ becomes very large (\red{$\sigma = 0.6$, blue curve}), a substantial fraction of samples exceed the critical threshold, thereby the spectrum more closely resembles the supercritical single-amplitude case: the UV enhancement is reduced, the peak becomes plateau-like, and the peak frequency shifts to lower values because shells are broader. These results illustrate that the fraction of large-amplitude shells strongly shapes the GW energy spectrum.

For comparison, we also perform a set of runs with a nonzero mean of $\mu = 0.4$ and the same $\sigma$ values, as shown in the right panel of Fig.~\ref{fig:sigma}. The results are qualitatively similar to the zero-mean case, except that the GW amplitude is generally larger due to the increased perturbation amplitudes. The spectrum for the single-amplitude case, $\mu = 0.4$ (black dashed line), corresponds to $\sigma = 0$, and generates the weakest GWs among the runs with $\sigma \ge 0$ due to the absence of larger-amplitude shells. \red{When the distribution contains a sufficiently large fraction of large-amplitude perturbations, nonperturbative contributions can distort the spectral shape, and further increasing $\sigma$ need not increase the GW amplitude monotonically.}

\subsection{Comparisons with SIGWs} \label{sec:compare_GW}

\begin{figure}[htbp]
    \centering
    \includegraphics[width=0.45\linewidth]{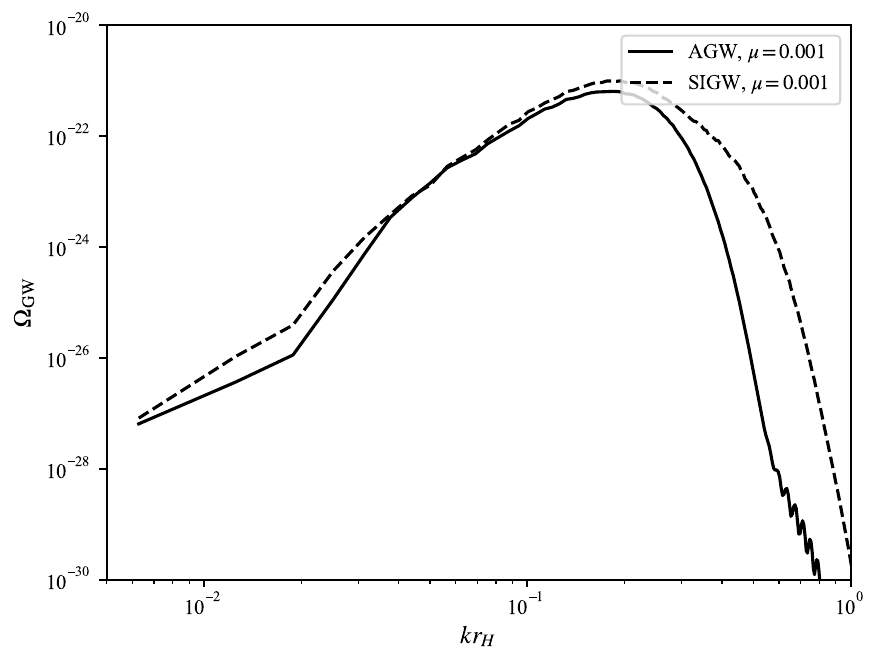}
    \qquad
    \includegraphics[width=0.45\linewidth]{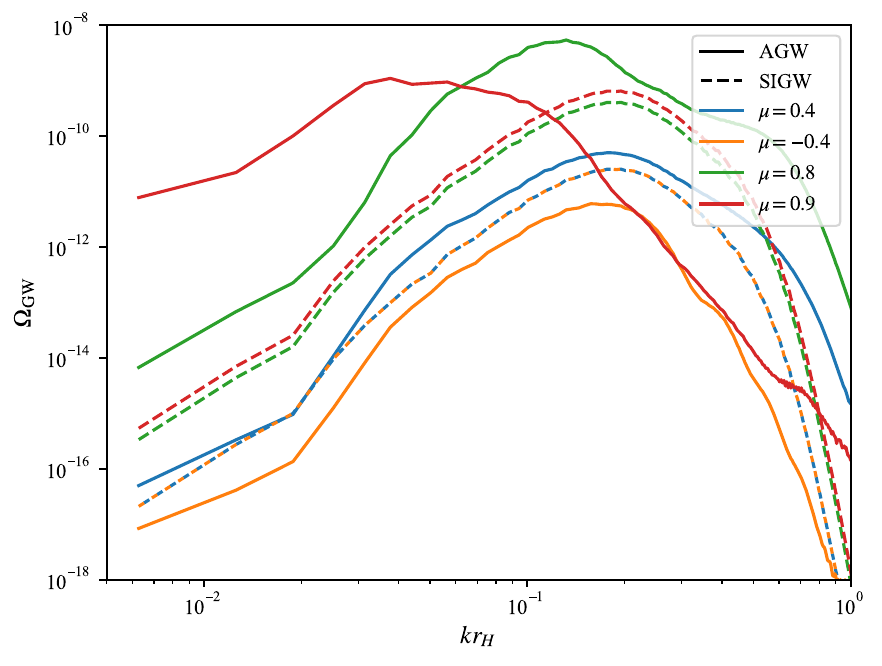}
    \caption{Comparison between the final energy spectra of Acoustic GWs (solid lines) and SIGWs (dashed lines). Left panel: small perturbation with $\mu = 0.001$. Right panel: various amplitudes of the curvature perturbations: $\mu = 0.4$ (blue), $\mu = -0.4$ (orange), $\mu = 0.8$ (green), and $\mu = 0.9$ (red). In the right panel, the blue and orange dashed lines overlap.}
    \label{fig:compare_GW}
\end{figure}

Our acoustic-GW calculation uses fully GR sound-shell profiles as initial conditions for 3D nonlinear hydrodynamic simulations while \red{neglecting metric backreaction on the fluid evolution}. SIGWs, on the other hand, are computed perturbatively (usually in Fourier space) and include contributions from both fluid motions and scalar metric perturbations to the given perturbative order. In Sec.~\ref {subsec:compare_SW}, we verified that, for sufficiently small perturbations, the fully GR and linear theory yield identical hydrodynamic evolutions; in that regime, SIGWs are expected to be larger because they, by definition, additionally include the metric contribution. When perturbations are large, however, nonlinear amplification of fluid motion can make acoustic GWs exceed the SIGW result. Below, we present a direct comparison computed from the same real-space curvature perturbations. SIGWs are strictly calculated using the recent method of real-space lattice simulations~\cite{Zeng:2025cer, Zeng:2025tno}: given a curvature perturbation configuration in real space, we evolve the equations of motion for scalar and tensor perturbations, thereby obtaining the energy spectrum of SIGWs.

For the initial condition used in the SIGW simulations, we embed $N_s = 50$ curvature peaks with a Gaussian profile~\eqref{eq:zeta} into a box of size $\tilde{L} = 1000$. We vary the amplitude parameter $\mu$ to study both small- and large-amplitude regimes. Acoustic GWs use the corresponding, fully GR sound-shell profiles extracted at $t_i = 200t_m$ as 3D initial data. This fluid configuration is actually the fully GR evolutionary outcome of the same curvature perturbations, ensuring a consistent comparison between acoustic GWs and SIGWs.

The left panel of Fig.~\ref{fig:compare_GW} compares final energy spectra of acoustic GWs (solid line) and SIGWs (dashed line) for the small-amplitude case $\mu = 0.001$. In this case, the hydrodynamic evolutions from the fully GR simulation and linear theory are identical, as we have already shown in the first panel of Fig.~\ref{fig:compare_SW}. The peak locations and overall shapes of the two spectra are similar, but the SIGW amplitude is slightly larger, consistent with the additional metric contribution in SIGW by definition. Therefore, for the small-perturbation case, our acoustic GW is simply the pure fluid motion part of SIGW, hence the name ``acoustic''. This also confirms that fluid motions are the dominant source of GWs, since the amplitude difference between the two is very small. 

The right panel shows the final energy spectra of acoustic GWs (solid lines) and SIGWs (dashed lines) for larger amplitudes of the curvature perturbations: $\mu = 0.4$ (blue), $\mu = -0.4$ (orange), $\mu = 0.8$ (green), and $\mu = 0.9$ (red). For the subcritical pairs ($\mu = 0.4$ and $\mu = -0.4$), the SIGW spectra are identical (the blue and orange dashed lines coincide). However, the acoustic GW spectrum for $\mu = 0.4$ ($\mu = -0.4$) is larger (smaller) than the SIGW, since the sound waves generated by curvature perturbations are stronger (weaker) than those predicted by linear theory due to nonperturbative effects. In the near-critical case ($\mu = 0.8$), acoustic GWs exceed SIGWs by roughly an order of magnitude, and the acoustic peak shifts to a lower frequency, consistent with nonlinear amplification in the UV and the broadening of the sound shell for the peak. For the supercritical case ($\mu = 0.9$), the acoustic peak shifts further to lower frequency and its amplitude decreases relative to $\mu = 0.8$ since only an underdense shell is produced in this case, which results in a weaker but broader sound shell. Nevertheless, the acoustic GW signal remains larger than SIGWs in our setup. Therefore, for the large-perturbation case, our acoustic GW, although still only accounts for the fluid motion contribution, goes beyond the perturbative prediction for SIGW due to the extra nonperturbative effect we consider for the initial profiles of sound shells.

These comparisons demonstrate that nonperturbative effects can substantially amplify the acoustic-GW channel and shift the peak frequency relative to SIGWs in the large-amplitude regime, while negative curvature perturbations can suppress the acoustic signal. A systematic exploration of the parameter space (including realistic amplitude distributions) is further left to future work.

\section{Conclusions and discussion} \label{sec:conclusions}

In this paper, accompanying previous studies~\cite{Zeng:2025law,Ning:2025ogq}, we have investigated the nonperturbative effect on the stochastic GW background sourced by primordial curvature perturbations, focusing specifically on the acoustic channel (fluid motions). The curvature perturbations are modeled as a collection of peaks, each of which evolves nonperturbatively to produce sound shells that subsequently collide and generate GWs. We have developed a hybrid numerical pipeline that (i) extracts fully GR, spherically symmetric sound-shell profiles from 1D Misner--Sharp simulations of individual curvature peaks, and (ii) embeds these nonlinear profiles into 3D lattice evolutions of relativistic hydrodynamics coupled to GWs. This strategy enables a controlled exploration of nonperturbative effects on GW production while avoiding the prohibitive cost of end-to-end 3D GR simulations.

\red{For simplified scenarios in which all curvature peaks share the same amplitude, we compared the numerical GW spectra with semi-analytical predictions from the sound shell model without rescaling either spectrum. The semi-analytical model captures the characteristic spectral shape and peak position, with the peak frequency closely tied to the comoving shell thickness through $k\simeq4/d$, but it does not reproduce the absolute numerical normalization exactly. The numerical-to-semi-analytical peak ratio depends on the physical case and on the simulation volume, rather than being described by a universal factor. At fixed $R_{*c}$, the ratio moves systematically closer to unity in the larger volume, consistent with finite-volume and finite-realization effects contributing to the mismatch; the present simulations do not establish an infinite-volume extrapolation. In the IR, all numerical spectra show a causal $k^3$ tail associated with the finite source duration. In the UV, nonlinear hydrodynamic interactions produce an enhancement relative to the linearized calculation. Removing these interactions makes the numerical spectrum follow the semi-analytical shape more closely, but leaves a residual normalization difference. We therefore do not attribute that residual to a single prefactor. The spectral shapes remain qualitatively similar for subcritical, negative-amplitude, and near-critical perturbations, whereas supercritical perturbations produce a plateau-like peak from the broader underdense shell. Independently of the absolute-normalization comparison, the peak amplitude follows the semi-analytical scaling $\Omega_{\rm GW}^{\rm peak}\propto R_{*c}^{-7}$ when the physical shell abundance is varied.}

We also studied the impact of an amplitude distribution of perturbations by sampling $\mu$ from a Gaussian distribution. The spectral shape depends on the distribution width $\sigma$, which controls the fraction of large-amplitude shells. As the distribution width increases, the overall amplitude increases, and the UV tail becomes slightly shallower. A very large value of $\sigma$ pushes a sizable fraction of samples above the collapse threshold, and the spectrum then resembles the supercritical single-amplitude case (reduced UV enhancement, plateau-like peak, lower peak frequency). The IR behavior remains universal and unaffected by the amplitude distribution.

Finally, we performed a direct comparison between acoustic GWs and SIGWs computed from the same real-space curvature perturbations. For small-amplitude perturbations, where the fully GR and linear theory yield identical hydrodynamic evolutions, SIGWs are slightly larger than acoustic GWs due to the additional metric contribution in SIGWs by definition. For large-amplitude perturbations, however, nonperturbative amplification of fluid motions can make acoustic GWs exceed SIGWs, with near-critical perturbations producing the largest enhancement (order-of-magnitude increases relative to SIGWs). The peak frequency of acoustic GWs also shifts to lower values due to nonlinear broadening of the sound shells. Negative curvature perturbations tend to suppress the acoustic GW signal. These results highlight the importance of nonlinear/nonperturbative effects for accurately predicting stochastic GW signals from primordial curvature perturbations.

The present study has several limitations that should be addressed in future work. The Gaussian curvature profile is a benchmark rather than a model-specific prediction. As demonstrated in Ref.~\cite{Ning:2025ogq}, changing the initial profile modifies the PBH threshold, sound-shell amplitude, thickness, and energy. The quantitative GW peak position, normalization, and detailed spectral shape are therefore profile dependent, even though sound-shell formation and outward propagation persist across the profiles investigated.

A further simplifying assumption is that all curvature peaks share the same characteristic scale $r_m$. For a broad primordial spectrum, peaks of different characteristic sizes would reenter the Hubble radius at different times and generate sound shells with different characteristic widths. Since the shell thickness controls the characteristic acoustic-GW frequency, a distribution of $r_m$ would produce a superposition of contributions over different scales, schematically
\begin{align}
    \Omega_{\rm GW}(k)
    \sim\int\mathrm{d}\ln r_m\,
    \mathcal W(r_m)\,
    \Omega_{\rm GW}(k;r_m)
    +\text{cross-correlations}.
\end{align}
Here $\mathcal W(r_m)$ and the cross-correlations must be derived from the underlying peak statistics rather than prescribed independently. A broad distribution of scales can therefore broaden or smooth the characteristic acoustic feature. A complete model-dependent prediction would require the joint distribution of peak height, scale, shape, and spatial position from a specified primordial spectrum. The present calculation should be understood as a benchmark response associated with a characteristic primordial scale rather than as a complete prediction for a broad primordial spectrum.

In addition, gravitational backreaction and the presence of any formed PBHs are neglected in the 3D lattice simulations. This is necessary for computational feasibility but omits potential GW contributions from scalar/vector metric perturbations and direct PBH-shell interactions. Finite spatial resolution induces numerical artifacts at the highest wavenumbers, and higher-resolution simulations are needed to further solidify the UV behavior. Our treatment of distributed $\mu$ uses a simple Gaussian sampling and is truncated to remain type-I; realistic primordial statistics, including non-Gaussianity, type-II perturbations, and spatial correlations, should be explored. Finally, standard SIGW calculations begin from a curvature power spectrum in momentum space, whereas our approach samples isolated peaks in real space. A direct connection between the two requires the peak statistics supplied by a specified primordial model and ultimately end-to-end 3D GR simulations.

In summary, we have developed a hybrid numerical approach to study the acoustic channel of GW production from primordial curvature perturbations. The spectra of acoustic GWs are governed by two physical length scales: the comoving shell thickness $d$ sets the characteristic frequency, while the mean comoving separation of sound shells $R_{*c}$ controls the overall amplitude. The IR and UV asymptotics are shaped by causality and nonlinear hydrodynamics, respectively. Although \red{gravitational backreaction on the fluid evolution is neglected}, our results clearly demonstrate that nonperturbative effects can substantially amplify the acoustic-GW channel and shift the peak frequency relative to SIGWs in the large-amplitude regime. The present work constitutes a first step toward a fully nonperturbative calculation of GWs from primordial curvature perturbations, while quantitative predictions for realistic primordial scenarios will require the extended numerical and analytical program outlined above.

\acknowledgments

This work is supported by the National Key Research and Development Program of China Grant No. 2021YFC2203004, No. 2021YFA0718304, and No. 2020YFC2201501, the National Natural Science Foundation of China Grants No. 12422502, No. 12547110, No.12588101, No. 12235019, and No. 12447101, and the Science Research Grants from the China Manned Space Project with No. CMSCSST-2021-B01 (supported by China Manned Space Program through its Space Application System). Z.-Y. Y. is supported by an appointment to the Young Scientist Training (YST) program at the APCTP through the Science and Technology Promotion Fund and Lottery Fund of the Korean Government. This was also supported by the Korean Local Governments-Gyeongsangbuk-do Province and Pohang City.



\bibliographystyle{JHEP}
\bibliography{biblio.bib}

\end{document}